\documentclass[twocolumn,english,pra,singlespace]{revtex4-1}
\usepackage[T1]{fontenc}
\usepackage[latin9]{inputenc}
\usepackage{color}
\usepackage{amsmath}
\usepackage{amssymb}
\usepackage{graphicx}

\makeatletter
\@ifundefined{textcolor}{}
{%
 \definecolor{BLACK}{gray}{0}
 \definecolor{WHITE}{gray}{1}
 \definecolor{RED}{rgb}{1,0,0}
 \definecolor{GREEN}{rgb}{0,1,0}
 \definecolor{BLUE}{rgb}{0,0,1}
 \definecolor{CYAN}{cmyk}{1,0,0,0}
 \definecolor{MAGENTA}{cmyk}{0,1,0,0}
 \definecolor{YELLOW}{cmyk}{0,0,1,0}
 }

\@ifundefined{definecolor}
 {\usepackage{color}}{}

\makeatother

\usepackage{babel}
\begin{document}

\title{Einstein-Podolsky-Rosen entanglement and steering in
two-well BEC ground states}

\author{Q. Y. He,$^{1}$, P. D. Drummond, $^{1}$, M. K. Olsen,$^{2}$ and
M. D. Reid$^{1}$}

\affiliation{$^{\text{1}}$Centre for Quantum Atom Optics, Swinburne University
of Technology, Melbourne, Australia\\
$^{\text{2}}$Centre for Quantum Atom Optics, University of Queensland,
Brisbane, Australia}
\begin{abstract}
We consider how to generate and detect Einstein-Podolsky-Rosen (EPR)
entanglement and the steering paradox between groups of atoms in two
separated potential wells in a Bose-Einstein condensate (BEC). We
present experimental criteria for this form of entanglement, and propose
experimental strategies for detecting entanglement using two or four
mode ground states. These approaches use spatial and/or internal modes.
We also present higher order criteria that act as signatures to detect\textcolor{black}{{}
the multiparticle entanglement present in this system. We point out
the difference between spatial entanglement using separated detectors,
and other types of entanglement that do n}ot require spatial separation.
The four-mode approach with two spatial and two internal modes results
in an entanglement signature with spatially separated detectors, conceptually
similar to the original EPR paradox. 
\end{abstract}
\maketitle

\section{Introduction\label{sec:Introduction}}

The Einstein-Podolsky-Rosen (EPR) paradox \cite{EPR paradox} established
a link between entanglement and nonlocality \cite{Bell} in quantum
mechanics. The extent to which entanglement can exist in spatially
separated macroscopic and massive systems is still essentially unknown.
Entanglement in optics however has been extensively studied and numerous
experiments have shown evidence for it \cite{entoptics,aspect,kwiatzeil,ou,eprcrit,rmp}.
An important distinction is that optical entanglement involves (nearly)
massless particles, and hence is a much less rigorous test of any
gravitational effects present.

Generation of EPR entanglement between two massive systems therefore
represents an important challenge. Such entanglement is a step in
the direction of fundamental tests of quantum mechanics, and is relevant
to the long term quest for understanding the relationship between
quantum theory and gravity. Ultimately, one would like to demonstrate
spatially entangled mass distributions, and this appears much more
promising for ultra-cold atoms than for room-temperature atoms. For
this reason, we focus on ultra-cold BEC environments here. This is
also relevant if BEC interferometry is to be useful to those areas
of quantum information and metrology where entanglement is known to
give an advantage \cite{Leeprl ,eprappli,spinsqwine,holburn,dowlent,fisher,naturepryentph}.
In this paper, we study strategies for generation of EPR entanglement
between Bose-Einstein condensates (BEC) confined to two spatially
separated potential wells.

Quantum correlations and EPR tests for Bose-Einstein condensates have
been suggested previously, with strategies involving molecular down-conversion
\cite{Moleccorrel} and four wave mixing interactions \cite{GardinerDeuar,olsenferris,coldmolebell},
among others. Early experiments measuring free-space correlations
demonstrated promising signatures of increased fluctuations associated
with entanglement \cite{Kasevich,Jin}, but were unable to conclusively
demonstrate entanglement or squeezing via reduced fluctuations, largely
due to measurement inefficiencies. This has improved with recent multi-channel
plate detection methods, but detection efficiency still remains an
issue \cite{westbrook}.\textcolor{red}{{} }Entanglement has also
been measured, very recently, for distinct but nearly spatially superimposed
modes \cite{neweprbec,eprenthiedel,science pairs ent} in an optical
lattice.

Here, we are motivated to study the two well case, in view of experiments
that have used this or similar systems to confirm both sub-shot noise
quantum correlations \cite{esteve}, and multiparticle entanglement
among a small group of atoms \cite{Gross2010,Philipp2010}. For much
larger numbers of atoms ($\sim40,000$), nearly quantum limited interferometry
has been recently verified \cite{Egorov}, showing that trapped atom
interferometry has the potential to reach mesoscopic sizes. There
have also been a number of previous theoretical studies \cite{bargill,bectheoryepr}
that outline different proposals and entanglement signatures.

The goal of this paper is to first clarify what it means to have an
EPR entanglement between groups of atoms in a BEC, and to outline
\textcolor{black}{a} strategy for achieving this goal. We define EPR
entanglement as being that entanglement existing between two spatially
separated systems, so that an EPR paradox can be realised. For EPR
entanglement to be claimed, three properties are to be evident \cite{rmp}: 
\begin{enumerate}
\item Two systems must be shown entangled through local measurements at
spatially distinct locations. 
\item The nature of the entanglement criterion should confirm an EPR paradox.
This requires measurement of sufficiently strong correlation between
the two systems, for two non-commuting {}``EPR'' observables such
as position/ momentum, conjugate spins, or quadrature phase amplitudes
\cite{eprcrit}. A generalised approach would allow other entanglement
measures, such as those for {}``EPR steering'' \cite{Schrodinger,hw-steering-1,hw2-steering-1,EPRsteering-1,hw-np-steering-1,asymmurray,multiqubits-1,loopholefreesteering}
which reveal an inconsistency between \textcolor{black}{EPR's }local
realism and the completeness of quantum mechanics using more general
measurement strategies. 
\item To fully justify EPR's no {}``spooky action-at-a-distance'' assumption
\cite{EPR paradox}, the measurement events should be causally separated
\cite{Bell,aspect,kwiatzeil}.
\end{enumerate}
For large groups of atoms, the task of detecting EPR entanglement
is much more feasible when the emphasis is on the EPR paradox itself,
rather than on the failure of Bell's local hidden variable model \cite{Bell}.
This leaves room for the possibility of confirming multiparticle entanglement,
a subject we touch on briefly in this paper. For spatially separated
systems, the detection of sufficient correlation of locally defined
EPR observables so that entanglement is confirmed \cite{Duan-simon-1,simon-1,proof for product form}
would represent an achievable first benchmark. This by itself is not\textcolor{black}{{}
direct e}vidence for the EPR paradox, or quantum steering, although
it is a necessary condition. The second step of confirming the paradox
has been carried out for photons \cite{rmp}, and also appears achievable
for atoms. The last step is probably the most difficult for atoms.
It would require either very fast measurements in one vacuum chamber,
or hybrid techniques involving two separated BECs with coupling via
atom-photon interfaces \cite{Interface}, in order to achieve causally
separated measurement.

There are many possible strategies for generation of spatial EPR entanglement.
Early experiments employed two photon cascades and, later, optical
parametric down conversion, to generate entangled photon pairs \cite{entoptics,aspect,kwiatzeil}.
Continuous variable EPR entanglement between two fields, in a so-called
{}``two-mode squeezed state'' \cite{caves sch}, was also generated
using parametric down conversion \cite{ou,eprcrit,eprparamp}. Such
entanglement gave evidence for an EPR paradox \cite{rmp}, although
true causal separation of measurement events was not demonstrated
in these experiments.

The paper is arranged as follows. In Section \ref{sec:Entanglement-Strategies}
we give a general introduction to the different possible entanglement
strategies. Section \ref{sec:Measurement-strategies} focuses on signatures
for demonstrating entanglement, pointing out the hierarchy of nonlocality
measures including EPR-steering \cite{hw-steering-1} and Bell's nonlocality
\cite{Bell}, as well as signatures for detecting multiparticle entanglement.
Section \ref{sec:generation-of-number-conserving} considers entanglement
preparation in a two-well system, modeled as two modes with boson
operators $a$ and $b$ \cite{Leeprl }. In this case, the S-wave
scattering \emph{intra-well }interactions, given by Hamiltonians $H=ga^{\dagger2}a^{2}$
and $H=gb^{\dagger2}b^{2}$, provides a local nonlinearity at each
well, while the coupling or tunneling \emph{inter-well} term, modeled
as $H=\kappa(a^{\dagger}b+ab^{\dagger})$ generates inter-well entanglement.
Here the intra- and inter-well interactions act \emph{simultaneously},
to enhance entanglement formation in the ground state. Section \ref{sec:EPR-entanglement:-Four}
treats a four-mode generalization of this, which has the advantage
that EPR-entanglement can be measured using atom counting at each
site, without the use of a local oscillator. Our conclusions are summarized
in Section \ref{sec:Summary}, with details given in the Appendices.
This paper is based on preliminary work presented in a Letter \cite{bectheoryepr}.
A second class of entanglement strategies using dynamical techniques
will be analyzed in a subsequent paper.

\section{Entanglement Strategies\label{sec:Entanglement-Strategies}}

\subsection{Prototype states for two-mode entanglement}

Suppose two spatially separated systems are describable as distinct
modes, represented by boson operators $a$ and $b$. There are two
prototype states that one can consider, that can give multiparticle
EPR entanglement. The first, which we ca\textcolor{black}{ll particle-pair
generation, is} currently the most widely known and used \cite{ou}.
We consider an entangled state with number correlations: 
\begin{equation}
|\psi\rangle_{II}=\sum_{n=0}^{\infty}c_{n}|n\rangle_{a}|n\rangle_{b}.\label{eq:sq2boson-1}
\end{equation}
 This type of two-mode squeezed state gives two-particle correlations
arising from a pair production process $H=\kappa a^{\dagger}b^{\dagger}+\kappa^{*}ab$
where $\langle ab^{\dagger}\rangle=0$ but $\langle ab\rangle\neq0$,
and the number difference is always squeezed \cite{number french,lane abs}.
These EPR states are formed in optics with parametric down conversion
\cite{eprcrit,rmp}, and similarly in nondegenerate four wave mixing
\cite{four wave mixing squeezing}. Since they are \textbf{not} number-conserving,
they are not typical of states formed in coupled two-well experiments,
although they have been generated in recent BEC experiments using
spin or mode-changing collisions \cite{neweprbec,eprenthiedel,science pairs ent}.

In this paper, we will focus on a second form \textcolor{black}{of
EPR entanglement, }which we call number conserving. This occurs, for
example, when fixed number states are input into a beam splitter:
$H=\kappa(a^{\dagger}b+ab^{\dagger})$, so that $\langle ab\rangle=0$
but $\langle ab^{\dagger}\rangle\neq0$. We consider an entangled
number-conserving state of form \cite{bec well ham cirac,gerdan,gorsav,wellbecmurr,supbec,carr}:
\begin{equation}
|\psi\rangle_{I}=\sum_{n=0}^{N}c_{n}|n\rangle_{a}|N-n\rangle_{b}.\label{eq:sq1-1}
\end{equation}
 This is the closest to the state prepared in some recent two-well
BEC experiments, where the total number is conserved \cite{esteve,Gross2010}.
We will examine how to unambiguously detect two-mode entanglement,
and EPR steering entanglement, for these states.

\subsection{Experimental strategies}

Before examining detailed solutions for an interacting BEC, it is
useful to summarize how two-mode number-conserving entanglement can
be generated, in schematic form. We consider how to generate entanglement
between two groups of atoms in separated potential wells in a BEC.
What is useful is a combination of \textbf{\emph{nonlinear local}}
interactions to generate a nonclassical squeezed state in each well
- together with a \textbf{\emph{nonlocal linear}} interaction to produce
the entanglement between two spatially distinct locations. In the
case of the BEC, the S-wave scattering can provide a nonlinear local
interaction, and quantum diffusion across a potential barrier acts
like a beam-splitter to provide the final nonlocal linear interaction.
Both effects occur at the same time in the schemes treated here, in
Sections \ref{sec:generation-of-number-conserving} and \ref{sec:EPR-entanglement:-Four}.

\begin{figure}
\begin{centering}
\includegraphics[width=0.5\columnwidth]{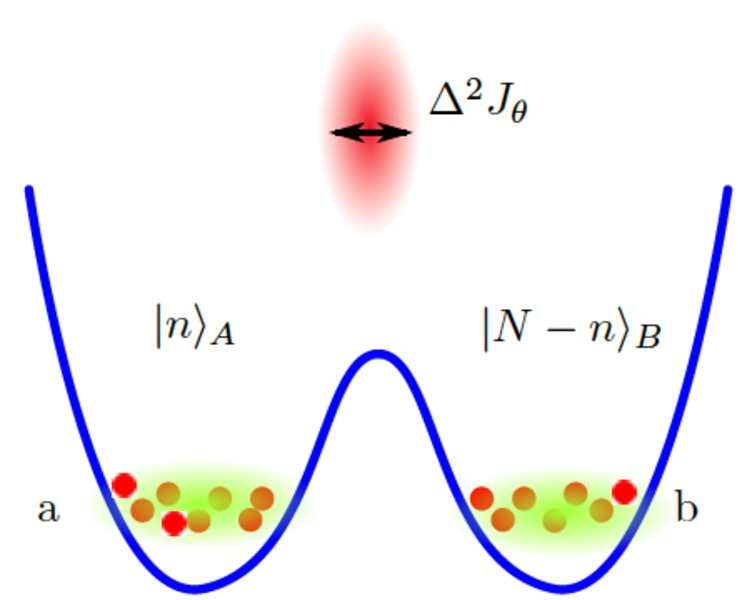} 
\par\end{centering}

\caption{(Color online) Two mode case: A double-well, one spin orientation
BEC. $a,$$b$ are operators for two modes at $A$ and $B$. The $a$
and $b$ are prepared with a two mode number difference squeezing
and an entanglement, by adiabatic cooling to the ground state. We
develop signatures to detect the inter-well entanglement, using inter-well
spin operators. \label{fig:system-two modes}}
\end{figure}

\begin{figure}[h]

\begin{centering}
\includegraphics[width=0.7\columnwidth]{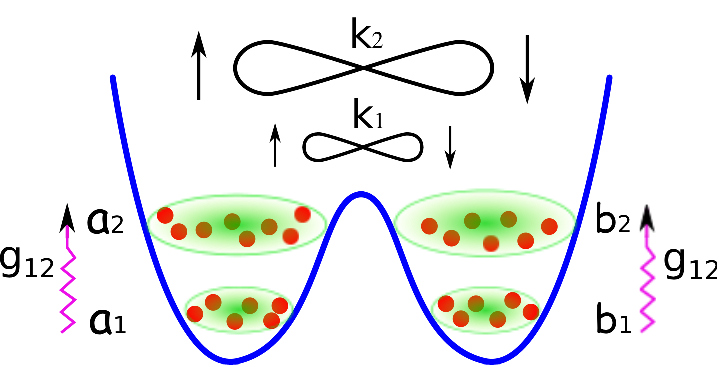} 
\par\end{centering}

\caption{(Color online) Four-mode case: A double-well, two spin orientation
BEC. We suppose the modes $a_{i}$ and $b_{i}$ are spatially separated.
Modes $a_{1}$, $a_{2}$ could be different spatial modes, or different
spin components of the same well. Pairs $a_{1}$, $b_{1}$ (and $a_{2}$,
$b_{2}$) can become entangled, due to the inter-well couplings. We
allow for the asymmetric case where pair $a_{2}$ and $b_{2}$ have
much greater numbers than $a_{1}$ and $b_{1}$ ($N_{2}\gg N_{1}$)
and also consider a case where modes $a_{2}$ and $b_{2}$ need not
be entangled ($\kappa_{2}=0$\textcolor{black}{). \label{fig:four-modes} }}
\end{figure}

We show in Section \ref{sec:generation-of-number-conserving} that
the entanglement generated for the two-well ground state with a fixed
number of atoms can translate to an EPR steering type of entanglement
\cite{hw-steering-1,multiqubits-1} (Fig. \ref{fig:system-two modes}).
For an \emph{actual }demonstration of this sort of EPR entanglement,
however, one must use signatures that involve local measurements,
for two spatially separated observers (often called Alice and Bob),
at sites $A$ and $B$. One can use local oscillator (LO) measurements
at each site, that provide phase shifts or their equivalent between
the measured and LO modes \cite{olsenferris,eprenthiedel}. \textcolor{black}{In
Section V, we propose an alternative though similar four-mode strategy,
as shown in Fig. \ref{fig:four-modes}. We summarise the two types
of }\textcolor{black}{\emph{gedanken-experiment}}\textcolor{black}{{}
as follows:}

\begin{itemize}
\item \textbf{Two-mode entanglement preparation then analysis}: the entangled
state is generated as the two-mode ground state in a double-well potential
(Fig. 1). Experimentally, this appears relatively simple, involving
evaporative cooling to the ground state in a single well followed
by an adiabatic ramping of an optical lattice to provide the central
potential barrier \cite{Bloch,esteve}. However, there are two levels
of experimental demonstration of the entanglement. The simplest involves
a nonlocal measurement that recombines the two modes, to demonstrate
an interwell entanglement. For demonstration of the EPR steering paradox,
however, strictly local measurements must be used. EPR steering entanglement
can be detected with a phase sensitive {}``local oscillator'' measurement
at each well, though this may represent an experimental challenge.
This strategy is discussed in Section \ref{sec:generation-of-number-conserving}. 
\item \textbf{Four-mode entanglement preparation then analysis}: we consider
four-mode states created through cooling in a double-well potential
with two spin states in each well (Fig. 2). Experimentally, this is
more complex, but an EPR steering entanglement can be demonstrated
using local Rabi rotations of the two spins of each well. This strategy
is discussed in Section \ref{sec:EPR-entanglement:-Four}. 
\end{itemize}

In both two and four mode cases, the basic idea is: 
\begin{enumerate}
\item Correlated ground state preparation, through evaporative cooling in
a potential well with linear coupling between wells.\emph{ }
\item Local Rabi rotation (in the four-mode case) to a superposition of
internal spins, thus\textcolor{black}{{} choosing an EPR measurement
angle.} In the two mode case, entanglement can be detected by nonlocal
rotation of the two spins.
\item Measurement, usually from absorption imaging, giving occupation numbers. 
\end{enumerate}

\section{Entanglement and EPR Steering criteria\label{sec:Measurement-strategies}}

In the original EPR proposal \cite{EPR paradox}, the paradox arose
from correlations between the positions and momenta of two particles
emitted from the same source. With optical or atomic Bose fields,
one can define the quadrature phase amplitudes of the modes, as $X_{A}=a^{\dagger}+a,$
and $Y_{A}=(a^{\dagger}-a)/i$, and similarly for mode $b$. These
have similar commutators to position and momentum in the particle
system. Detection of sufficient correlation between the quadratures
will signify entanglement \cite{Duan-simon-1,simon-1}, and the EPR
paradox \cite{eprcrit}, as analysed recently for atoms by Gross et
al \cite{eprenthiedel}. 

We find that the common approach of detecting the EPR correlation
as a reduced variance \cite{Duan-simon-1,eprcrit} is not so useful
for the number conserving entangled states (\ref{eq:sq1-1}). Instead,
we adapt the criteria proposed by Hillery and Zubairy \cite{hillzub},
and Cavalcanti et al \cite{CFRD,multiqubits-1,zubhighmomentbell,bellcfrdnew,bellcfrdpra}.
Like most practical criteria to date, these methods are sufficient,
but not necessary, for the detection of entanglement. The limitations
of measures of entanglement based on purity have been pointed out
recently by Chianca and Olsen \cite{ch olsen ent measurewell }.

\subsection{Two-mode Hillery-Zubairy entanglement criterion}

Two subsystems $A$ and $B$ are said to be entangled if the density
operator $\rho$ for the composite system cannot be expressed as a
mixture of product states ie.
\begin{equation}
\rho=\sum_{R}P_{R}\rho_{A}^{R}\rho_{B}^{R}\label{eq:rhosep}
\end{equation}
fails, where $\sum_{R}P_{R}=1$, and $\rho_{A/B}^{R}$ is a density
operator for $A/B$. 

Consider where systems are single field modes with boson operators
$a$ and $b$ respectively. Hillery and Zubairy (HZ) showed that the
two modes $a$ and $b$ are entangled if \cite{hillzub} 
\begin{equation}
|\langle a^{m}\left(b^{\dagger}\right)^{n}\rangle|^{2}>\langle\left(a^{\dagger}\right)^{m}a^{m}\left(b^{\dagger}\right)^{n}b^{n}\rangle.\label{eq:hilzub2-1-1-1}
\end{equation}
All separable states (defined as those for which (\ref{eq:rhosep})
holds) satsify $|\langle a^{m}\left(b^{\dagger}\right)^{n}\rangle|^{2}\leq\langle\left(a^{\dagger}\right)^{m}a^{m}\left(b^{\dagger}\right)^{n}b^{n}\rangle$. 

In Ref \cite{bectheoryepr}, we suggested how to rewrite criterion
(\ref{eq:hilzub2-1-1-1}) for $m=n$. For any nonhermitian operator
$Z$, we consider the generalized variance, which must be nonnegative:
\textcolor{black}{{} 
\begin{equation}
\Delta^{2}Z\equiv\langle\left(Z^{\dagger}-\langle Z^{\dagger}\rangle\right)\left(Z-\langle Z\rangle\right)\rangle=\langle Z^{\dagger}Z\rangle-\langle Z^{\dagger}\rangle\langle Z\rangle\geq0\,.
\end{equation}
 Defining $Z=a^{m}\left(b^{\dagger}\right)^{m}$, we find it is always
true (for any state) that 
\begin{eqnarray}
|\langle a^{m}b^{\dagger m}\rangle|^{2}-\langle a^{\dagger m}a^{m}b^{\dagger m}b^{m}\rangle & \leq & \langle a^{\dagger m}a^{m}([b^{m}b^{\dagger m},b^{\dagger m}b^{m}])\rangle.\nonumber \\
\label{eq:momenthz}
\end{eqnarray}
 Thus, the HZ criterion (\ref{eq:hilzub2-1-1-1}) confirms entanglement
if: 
\begin{eqnarray}
0\leq E_{HZ}^{(m)} & = & 1+\frac{\langle a^{\dagger m}a^{m}b^{\dagger m}b^{m}\rangle-|\langle a^{m}b^{\dagger m}\rangle|^{2}}{\langle a^{\dagger m}a^{m}(b^{m}b^{\dagger m}-b^{\dagger m}b^{m})\rangle}<1.\nonumber \\
\label{eq:HZentcritm}
\end{eqnarray}
 It is also possible to derive a criterion using the commutators for
mode $a$. Hence the HZ entanglement criterion (\ref{eq:HZentcritm})
is best written with the optimal choice of denominator, corresponding
to the minimum of $\langle a^{\dagger m}a^{m}(b^{m}b^{\dagger m}-b^{\dagger m}b^{m})\}$
or $\langle b^{\dagger m}b^{m}(a^{m}a^{\dagger m}-a^{\dagger m}a^{m})\rangle$.}

The\emph{ first order} ($m=n=1$) HZ criterion for entanglement becomes\textcolor{black}{{}
\begin{eqnarray}
0<E_{HZ}^{(1)} & = & 1+\frac{\langle a^{\dagger}ab^{\dagger}b\rangle-|\langle ab^{\dagger}\rangle|^{2}}{min\{\langle a^{\dagger}a\rangle,\langle b^{\dagger}b\rangle\}}<1.\label{eq:hz2}
\end{eqnarray}
 }

\subsection{Multiparticle entanglement criterion}

The \emph{second order} HZ entanglement criterion is obtained by using
the power $m=2$ with the identity $[b^{2}b^{\dagger2},b^{\dagger2}b^{2}]=4b^{\dagger}b+2$.
Entanglement is then observed if\textcolor{black}{{} 
\begin{eqnarray}
0\leq E_{HZ}^{(2)} & = & 1+\frac{\langle a^{\dagger2}a^{2}b^{\dagger2}b^{2}\rangle-|\langle a^{2}b^{\dagger2}\rangle|^{2}}{\langle a^{\dagger2}a^{2}(4b^{\dagger}b+2)\rangle}<1.\nonumber \\
\label{eq:hzsecond-1}
\end{eqnarray}
}We now show that the higher order HZ entanglement criterion (\ref{eq:HZentcritm})
with $m>1$ enables detection of multi-particle entanglement. The
criterion (\ref{eq:hzsecond-1}) can only be satsified if there exists
a nonzero probability that the system is in an entangled superposition
state of the form
\begin{equation}
|\psi\rangle=c|n_{A}\rangle|n_{B}\rangle+d|n_{A}+m\rangle|n_{B}'\rangle+\sum_{n,l}c_{nl}|n\rangle|l\rangle\label{eq:entsupm}
\end{equation}
\textcolor{black}{(or that obtained by interchanging the states of
$A$ and $B$) where the amplitudes $c,d\neq0$ but $c_{nl}$ are
unspecified. Here $|n_{A}\rangle|n_{B}\rangle$ is the product number
state with $n_{A}$ particles in $A$ and $n_{B}$ particles in $B$.}

\textcolor{black}{\emph{Proof: }}\textcolor{black}{Any composite system
$A$/ $B$ can be described by a density matrix $\rho=\sum_{R}P_{R}\rho_{sep}^{R}+\sum_{R'}P_{R'}\rho_{ent}^{R'}$,
where $\rho_{sep}^{R}$ and $\rho_{ent}^{R'}$ represent pure separable
and entangled states respectively. The higher-order HZ entanglement
measure (\ref{eq:hilzub2-1-1-1}) with $m=n$ can therefore be written
as a ratio 
\begin{eqnarray}
R=\frac{|\langle a^{m}\left(b^{\dagger}\right)^{m}\rangle|^{2}}{\langle\left(a^{\dagger}\right)^{m}a^{m}\left(b^{\dagger}\right)^{m}b^{m}\rangle}\label{eq:R-1}
\end{eqnarray}
 where 
\begin{eqnarray*}
\langle a^{m}\left(b^{\dagger}\right)^{m}\rangle & = & \sum_{R}P_{R}\langle a^{m}\left(b^{\dagger}\right)^{m}\rangle_{R}+\sum_{R'}P_{R'}\langle a^{m}\left(b^{\dagger}\right)^{m}\rangle_{R'}
\end{eqnarray*}
 and 
\begin{eqnarray*}
\langle\left(a^{\dagger}\right)^{m}a^{m}\left(b^{\dagger}\right)^{m}b^{m}\rangle & = & \sum_{R}P_{R}\langle\left(a^{\dagger}\right)^{m}a^{m}\left(b^{\dagger}\right)^{m}b^{m}\rangle_{R}\\
 &  & +\sum_{R'}P_{R'}\langle\left(a^{\dagger}\right)^{m}a^{m}\left(b^{\dagger}\right)^{m}b^{m}\rangle_{R'}
\end{eqnarray*}
 Here $\langle O\rangle_{R}$ represents the expectation value of
$O$ for state $\rho_{R}$. Since for a separable state, $R\leq1$,
we can see that if $\sum_{R'}P_{R'}\langle a^{m}\left(b^{\dagger}\right)^{m}\rangle_{R'}=0$,
it is always the case that $\rho$ predicts $R\leq1$. In short, the
higher order entanglement, $E_{HZ}^{(m)}<1$, cannot be achieved unless}
there is a nonzero probability $P_{R'}$ for a pure entangled state
$\rho_{ent}^{R'}$ for which $\langle a^{m}\left(b^{\dagger}\right)^{m}\rangle\neq0$.
Expanding $\rho_{ent}^{R'}$ in terms of the number state basis $|n_{A}\rangle|n_{B}\rangle$
where yields $\rho_{ent}^{R'}=|\psi\rangle\langle\psi|$, where 
\begin{equation}
|\psi\rangle=\sum_{n,l}c_{nl}|n\rangle_{A}|l\rangle_{B}.\label{eq:sup-1}
\end{equation}
 If only adjacent number states $|n\rangle$, $|n+1\rangle$ have
nonzero amplitude $c_{nl}$, the $\langle a^{m}(b^{\dagger})^{m}\rangle=0$
where $m>1$. Hence, the superposition (\ref{eq:sup-1}) necessarily
includes number states separated by $ $$m$.

\subsection{Two-mode EPR-steering criterion}

Nonlocality can be revealed using criteria similar to (\ref{eq:hilzub2-1-1-1}).
Entanglement itself does not imply an EPR-steering paradox \cite{EPR paradox,eprcrit,hw-steering-1,EPRsteering-1}
nor violation of local hidden variable theories (Bell's theorem) \cite{CFRD,zubhighmomentbell,bellcfrdnew,bellcfrdpra,HDRspin,multiqubits-1},
which are seen as stronger forms of entanglement. In this paper, we
consider two sites only, and focus on the entanglement and EPR-steering
cases, since it has been shown that violation of the moment Bell inequality
derived in Ref \cite{CFRD} requires three or more sites \cite{bellcfrdnew}.

The EPR paradox was discussed by Schrodinger \cite{Schrodinger},
who introduced the notion of {}``steering'' as an apparent action-at-a-distance.
Criteria for {}``steering'' can be developed using the asymmetric
local hidden state separable model of Wiseman and co-workers \cite{hw-steering-1}.
Violation of this model reveals inconsistency of EPR's asymmetric
local realism with the completeness of quantum mechanics, and thus
may be thought of as a generalized EPR paradox \cite{hw-steering-1,hw2-steering-1,EPRsteering-1,rmp}.
The EPR paradox-steering nonlocality has been realised experimentally
in loop-hole free and high efficiency scenarios for optical qubits
\cite{loopholefreesteering} and Gaussian states \cite{ou,rmp}. 

An EPR-steering nonlocality\textbf{ }is detected if 
\begin{eqnarray}
|\langle a^{m}b^{\dagger n}\rangle|^{2} & > & \langle a^{\dagger m}a^{m}(\frac{b^{\dagger n}b^{n}+b^{n}b^{\dagger n}}{2})\rangle.\label{eq:steerhzone}
\end{eqnarray}
 The proof follows from straightforward application of methods is
given in \cite{multiqubits-1} which derived this EPR steering criterion
for $m=n=1$. This criterion can also be rewritten in terms of the
HZ entanglement parameter (\ref{eq:hz2}), so that \textcolor{black}{\emph{EPR-steering
entanglement}}\textcolor{black}{{} is confirmed if:}

\textcolor{black}{
\begin{eqnarray}
0\leq E_{HZ}^{(m)} & = & 1+\frac{\langle a^{\dagger m}a^{m}b^{\dagger m}b^{m}\rangle-|\langle a^{m}b^{\dagger m}\rangle|^{2}}{\langle a^{\dagger m}a^{m}(b^{m}b^{\dagger m}-b^{\dagger m}b^{m})\rangle}<\frac{1}{2}.\nonumber \\
\label{eq:eprsteerhz}
\end{eqnarray}
}

We note that the moments of type $\langle a^{m}b^{\dagger n}\rangle$
are in principle measured as a linear combination of moments of the
Hermitian observables, $X_{A}$ and $P_{A}$, and $X_{B}$ and $P_{B}$.

\subsection{Two-mode spin entanglement and EPR steering criteria}

It is convenient to quantify entanglement using spin-operator methods,
the advantage being that for BEC two-well systems, the variances of
Schwinger spins have been measured in experiment \cite{esteve}. Hillery
and Zubairy \cite{hillzub} have written the first order criterion
(\ref{eq:hilzub2-1-1-1}) in terms of the variances of inter-well
Schwinger spins, defined as: 
\begin{eqnarray}
J_{AB}^{X} & = & \left(a^{\dagger}b+ab^{\dagger}\right)/2\nonumber \\
J_{AB}^{Y} & = & \left(a^{\dagger}b-ab^{\dagger}\right)/(2i)\nonumber \\
J_{AB}^{Z} & = & \left(a^{\dagger}a-b^{\dagger}b\right)/2\nonumber \\
J_{AB}^{2} & = & \hat{N}_{AB}(\hat{N}_{AB}+2)/4\nonumber \\
\hat{N}_{AB} & = & a^{\dagger}a+b^{\dagger}b\,.\label{eq:schspinab}
\end{eqnarray}
 \textcolor{black}{Where the outcomes for $\hat{N}_{AB}$ are fixed
at $N$, the spin is fixed as $J=N/2$. The HZ entanglement criterion
given by Eq. (\ref{eq:hz2}) for $m=n=1$ can then be rewritten as:
\begin{eqnarray}
0<E_{HZ} & = & \frac{\left(\Delta J_{AB}^{X}\right)^{2}+\left(\Delta J_{AB}^{Y}\right)^{2}}{\langle\hat{N}_{AB}\rangle/2}<1.\label{eq:HZ-spin version-1-1-2}
\end{eqnarray}
 }We recall from (\ref{eq:eprsteerhz}) that EPR steering is observed
if \textcolor{black}{
\begin{eqnarray}
0<E_{HZ} &  & <1/2.\label{eq:HZ-spin version-1-1-2-1}
\end{eqnarray}
 }It should be noted here that this type of spin-operator variance
has been measured experimentally \cite{esteve} by observing the interference
between the two modes, on expanding the atomic clouds after turning
the traps off. However, as we discuss later, this strategy cannot
be readily interpreted in the EPR sense, due to the lack of separation
during measurement.

\textcolor{black}{The best entanglement (for a fixed number of atoms
$N$) as measured by (\ref{eq:HZ-spin version-1-1-2}) is given when
the sum of the two variances of $J_{AB}^{X}$ and $J_{AB}^{Y}$ is
minimized. This sum can }\textcolor{black}{\emph{never}}\textcolor{black}{{}
be zero, meaning that the ideal entanglement of $E_{HZ}=0$ cannot
be reached, because the spins $J_{AB}^{X}$ and $J_{AB}^{Y}$ do not
commute. However, the sum becomes asymptotically small for large $N$,
in which case large noise appears in the third spin $J_{AB}^{Z}$.
The lower bound for the sum of the two variances has been obtained
by \cite{cj}: 
\begin{equation}
\frac{\left(\Delta J_{AB}^{X}\right)^{2}+\left(\Delta J_{AB}^{Y}\right)^{2}}{J}\geq C_{J}/J\label{eq:cj-1}
\end{equation}
 where the coefficients $C_{J}$ are given in that reference. The
reduction of the sum $\left(\Delta J_{AB}^{X}\right)^{2}+\left(\Delta J_{AB}^{Y}\right)^{2}$
below the standard quantum limit (given by $J=\langle\hat{N}_{AB}\rangle/2$)
is referred to as {}``planar squeezing'', and represents the onset
of HZ entanglement. }

\textcolor{black}{Inequalities of the type (\ref{eq:cj-1}) are useful
for inferring multiparticle entanglement. The level of entanglement
as measured by $E_{HZ}$ can give information about how many atoms
are involved in the entangled state. Since a large spin $J$ can only
be obtained where the number of atoms $N$ is large, very small squeezing
necessarily implies an entangled state with a large mean $\langle N\rangle$.
This approach was developed by Sorenson and Molmer \cite{sorsolm},
who explained how to infer a multiparticle entanglement from the level
of reduction in the {}``spin squeezing'' variance of $J_{Z}$ \cite{Gross2010,Philipp2010}.}

\subsection{Four-mode spin EPR entanglement criteria}

A true EPR experiment would involve coherent combination of second
fields or condensates at each site, as depicted schematically in Fig.
\ref{fig:four-modes}. To observe true EPR entanglement between sites
$A,B$, a useful procedure is to use two modes per EPR site. Local
intra-well spin measurements are defined: for well $A$, 
\begin{eqnarray}
J_{A}^{X} & = & \left(a_{1}^{\dagger}a_{2}+a_{2}^{\dagger}a_{1}\right)/2,\nonumber \\
J_{A}^{Y} & = & \left(a_{1}^{\dagger}a_{2}-a_{2}^{\dagger}a_{1}\right)/(2i),\nonumber \\
J_{A}^{Z} & = & \left(a_{2}^{\dagger}a_{2}-a_{1}^{\dagger}a_{1}\right)/2,\nonumber \\
\hat{N}_{A} & = & a_{2}^{\dagger}a_{2}+a_{1}^{\dagger}a_{1}.\label{eq:Intrawell-spin}
\end{eqnarray}
 Here $a_{1,2}$ are mode operators for different components of the
same \textcolor{black}{site}, typically different spatial modes or
different nuclear spins at each site. We will also introduce the notation
for the corresponding raising and lowering spin operators, $J_{A}^{\pm}=J_{A}^{X}+iJ_{A}^{Y}$.
Similar spin operators are defined for site $B$.\textcolor{red}{{}
}This defines complementary observables that are locally measurable
at each site, using Rabi rotations and number-difference measurements.
Calculations of spin correlations at two sites can be carried out
most simply on imaging on a micron scale, then dividing the imaged
atoms into two halves for measurement purposes. A more sophisticated
method is to add a time-dependent external potential to divide the
condensate into two widely separated parts. While this gives results
that depend on the potential, it provides a physical separation between
the sites.

Having defined local spin operators, we now need to consider a suitable
EPR entanglement measure. Previous authors have derived HZ-type entanglement
and EPR steering criteria that are expressed in terms of these effective
local spin operators \cite{hillspinpapers,spinhillery,mulithillery,HDRspin}.
Entanglement is confirmed if 
\begin{eqnarray}
\mid\langle J_{A}^{+}J_{B}^{-}\rangle\mid^{2} & > & \langle J_{A}^{+}J_{A}^{-}J_{B}^{+}J_{B}^{-}\rangle.\label{eq:entspinhilzub}
\end{eqnarray}
This inequality uses operators which are measurable locally using
Rabi rotations and number measurements \cite{Gross2010}. Criteria
involving higher moments are also possible, but are not examined here.
As for the original HZ criterion, the spin criterion can be rewritten
using the procedure outlined in \cite{bectheoryepr}. If we define
$Z=J_{A}^{+}J_{B}^{-}$, then we can easily show that $\Delta^{2}\left(J_{A}^{+}J_{B}^{-}\right)=\langle J_{A}^{+}J_{A}^{-}J_{B}^{+}J_{B}^{-}\rangle-\langle[J_{A}^{+},J_{A}^{-}]J_{B}^{+}J_{B}^{-}\rangle-|\langle J_{A}^{+}J_{B}^{-}\rangle|^{2}\geq0.$
Thus, 
\begin{eqnarray}
{\color{black}|\langle J_{A}^{+}J_{B}^{-}\rangle|^{2}-\langle J_{A}^{+}J_{A}^{-}J_{B}^{+}J_{B}^{-}\rangle} & {\color{black}\leq} & {\color{black}\langle[J_{A}^{-},J_{A}^{+}]J_{B}^{+}J_{B}^{-}\rangle}\nonumber \\
{\color{black}} & {\color{black}=} & {\color{black}2\langle J_{A}^{Z}J_{B}^{+}J_{B}^{-}\rangle.}\label{eq:3}
\end{eqnarray}
 Similarly, defining $Z^{\dagger}=J_{A}^{-}J_{B}^{+}$, one can show
that 
\begin{eqnarray}
|\langle J_{A}^{+}J_{B}^{-}\rangle|^{2}-\langle J_{A}^{+}J_{A}^{-}J_{B}^{+}J_{B}^{-}\rangle & \leq & 2\langle J_{A}^{+}J_{A}^{-}J_{B}^{Z}\rangle.\label{eq:spincal}
\end{eqnarray}
 The spin entanglement criterion (\ref{eq:entspinhilzub}) becomes
\begin{eqnarray}
E_{HZ}^{spin\,(1)} & \text{=} & \frac{\Delta^{2}\left(J_{A}^{+}J_{B}^{-}\right)}{min[2\langle J_{A}^{Z}J_{B}^{+}J_{B}^{-}\rangle,2\langle J_{A}^{+}J_{A}^{-}J_{B}^{Z}\rangle]}<1\,\label{eq:hzspinvarcrit}
\end{eqnarray}
 i.e. HZ-type spin entanglement is verified if $E_{HZ}^{spin\,(1)}<1$.

We have derived the spin EPR steering inequalities based on (\ref{eq:entspinhilzub})
in a previous paper \cite{HDRspin}. EPR steering is detected if 
\begin{eqnarray}
|\langle J_{A}^{+}J_{B}^{-}\rangle|^{2} & > & \langle[(J_{A})^{2}-(J_{A}^{Z})^{2}\pm J_{A}^{Z}]\nonumber \\
 &  & \ \ \times[(J_{B})^{2}-(J_{B}^{Z})^{2}]\rangle,\label{eq:EPRsteering-1}
\end{eqnarray}
 which can be rewritten as\textcolor{black}{{} 
\begin{eqnarray}
0\leq E_{HZ}^{spin\,(1)} & = & 1+\frac{\langle J_{A}^{+}J_{A}^{-}J_{B}^{+}J_{B}^{-}\rangle-|\langle J_{A}^{+}J_{B}^{-}\rangle|^{2}}{min[2\langle J_{A}^{Z}J_{B}^{+}J_{B}^{-}\rangle,2\langle J_{A}^{+}J_{A}^{-}J_{B}^{Z}\rangle]}<\frac{1}{2}.\nonumber \\
\label{eq:spinhzeprcritvar}
\end{eqnarray}
}

We note the spin moments of Eqs (\ref{eq:hzspinvarcrit}) and (\ref{eq:spinhzeprcritvar})
are actually measured via the $x$ and $y$ spin components, for example,
using the expansion: 
\begin{eqnarray}
\langle J_{A}^{+}J_{B}^{-}\rangle= & \langle J_{A}^{X}J_{B}^{X}- & iJ_{A}^{X}J_{B}^{Y}+J_{A}^{Y}J_{B}^{X}+J_{A}^{Y}J_{B}^{Y}\rangle.\nonumber \\
\label{eq:10}
\end{eqnarray}

\section{generation of two-mode entanglement\label{sec:generation-of-number-conserving} }

We next turn to physical means to generate and measure entanglement
and EPR-steering in two-mode physical systems. We focus here on the
\emph{gedanken}-experiment of Fig \ref{fig:system-two modes}, with
explicit spatial separation of the two modes.

\subsection{\textcolor{black}{Linear beam splitter with fixed number input states}}

Possibly the simplest number-conserving entangled state is obtained
with a number-squeezed input, together with a beam splitter interaction
\begin{equation}
H=\kappa a^{\dagger}b+\kappa^{*}ab^{\dagger}\,,\label{eq:hambs}
\end{equation}
 which models the exchange of atoms that can take place between wells.

\begin{figure}[h]

\begin{centering}
\includegraphics[width=0.5\columnwidth]{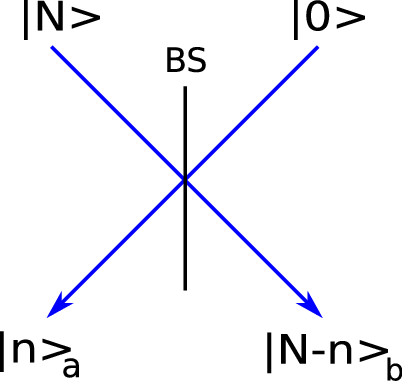} 
\par\end{centering}

\caption{(Color online) A Fock number state $|N\rangle$ incident on a beam
splitter produces an entangled state (\ref{eq:sq1-1}).\label{fig:single-fock-diagram}}
\end{figure}

On defining output ($a$, $b$), input ($a_{in}$) and vacuum ($a_{v}$)
input modes (Fig. \ref{fig:single-fock-diagram}, \ref{fig:double-Fock-diagram}),
one can write the beam splitter transformation as 
\begin{eqnarray}
a & = & (a_{in}+a_{v})/\sqrt{2}\label{eq:beamsplitter-1}\\
b & = & (a_{in}-a_{v})/\sqrt{2}.\nonumber 
\end{eqnarray}

\subsubsection{Single number-state input}

\begin{figure}[h]
\begin{centering}
\textcolor{blue}{\includegraphics[width=0.9\columnwidth]{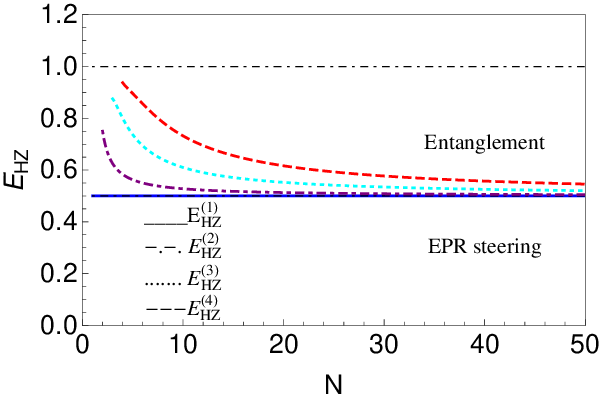}} 
\par\end{centering}

\caption{(Color online) A single Fock number state with beamsplitter is entangled
($E_{HZ}<1$) by the HZ entanglement criteria, Eq. (\ref{eq:HZentcritm}).
Higher order entanglement is indicated by the dashed lines. The correlation
does not confirm EPR-steering entanglement from Eq. (\ref{eq:eprsteerhz}),
which requires $E_{HZ}<0.5$.\label{fig:single-fock}}
\end{figure}

We first consider the simplest case of $N$ atoms input to one port
of the beam splitter (Fig. \ref{fig:single-fock-diagram}). This is
equivalent to the linear interferometer case \cite{Gross2010} in
which a fixed number of atoms are initially in one BEC well. These
are then redistributed between wells via a number conserving mechanism.
Using (\ref{eq:beamsplitter-1}), the final state is number-conserving
(\ref{eq:sq1-1}): 
\begin{eqnarray}
|out\rangle & = & \sum_{n=0}^{N}c_{n}|n\rangle_{a}|N-n\rangle_{b}\,,\label{eq:bsoutputstate}
\end{eqnarray}
 where $c_{n}=\sqrt{N!}/\sqrt{2^{N}n!(N-n)!}$. This state (\ref{eq:bsoutputstate})
is entangled for all $N$. The entanglement can be detected using
the Hillery and Zubairy entanglement measure (\ref{eq:HZentcritm})\textcolor{black}{.
The superposition (\ref{eq:bsoutputstate}) clearly involves up to
$N$ particles. This multiparticle entanglement can be detected using
the higher order entanglement $E_{HZ}^{(n)}$ criteria (\ref{eq:HZentcritm}).
Higher order (up to $N$-th) entanglement becomes evident in (Fig.
\ref{fig:single-fock}).}

This linear beam splitter method generates a relatively small degree
of entanglement, however, (Fig. \ref{fig:single-fock}), and will
later be compared with the much more significant entanglement obtainable
using nonlinear BEC interactions.

\subsubsection{Double number state input}

\begin{figure}[h]

\begin{centering}
\includegraphics[width=0.5\columnwidth]{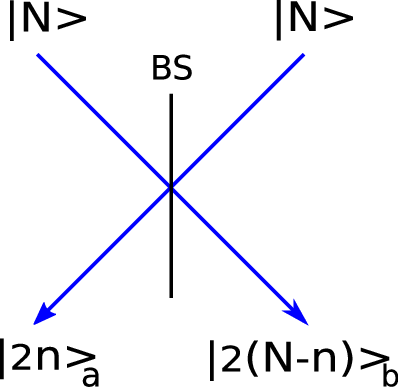} 
\par\end{centering}

\caption{(Color online) A double Fock number state incident on a beam splitter
also produces a number-conserving entangled state. \label{fig:double-Fock-diagram}}
\end{figure}

We next consider a double Fock number state $|N\rangle|N\rangle$
incident on a beam splitter (Fig. \ref{fig:double-Fock-diagram}),
as a model for the case where there is initially a fixed, equal number
of atoms in each well.

The output state after an exchange between the wells is 
\begin{eqnarray}
|out\rangle & = & \sum_{n=0}^{N}c_{n}|2n\rangle_{a}|2(N-n)\rangle_{b}\label{eq:outputsecondbs}
\end{eqnarray}
 where $c_{n}=(-1)^{N-n}\sqrt{(2n)!}\sqrt{\left(2(N-n)\right)!}/\left[2^{N}n!(N-n)!\right]$.
In this case, entanglement is again present for all $N$, but cannot
be detected via the first order entanglement criterion (\ref{eq:hz2}).

\begin{figure}[h]

\begin{centering}
\textcolor{blue}{\includegraphics[width=0.9\columnwidth]{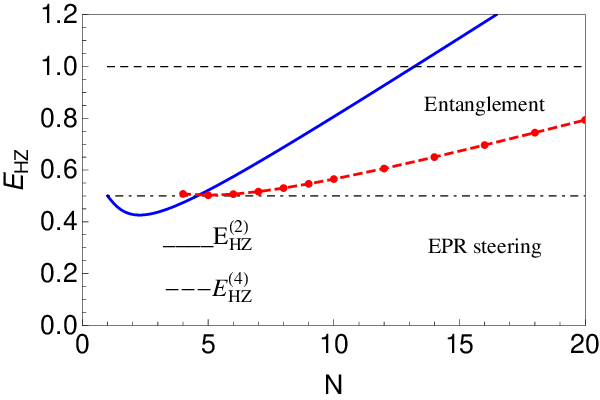}} 
\par\end{centering}

\caption{(Color online) HZ entanglement criterion using a double Fock number
state and beam splitter. The graph shows the criterion (\ref{eq:HZentcritm})
for $m=n=2$ (solid blue line), and $m=n=4$ (red dashed line). EPR-steering
is observable with $m=n=2$ and $N<5$.\label{fig:double-Fock}}
\end{figure}

Entanglement can however be detected via the second order HZ entanglement
criterion Eq. (\ref{eq:hzsecond-1}\textcolor{black}{), which indicates
an entanglement}\textcolor{black}{\emph{ }}\textcolor{black}{involving
a superposition of number states different by two particles (proved
in Section III.B). The fourth-order entanglement $E^{(4)}$ is also
evident, indicating superpositions involving states separated by four
particles. The entanglement measure $E^{(2)}$ is sufficiently strong
that EPR steering can also be confirmed via Eq. (\ref{eq:eprsteerhz})
with }$m=n=2$, as shown in Fig. \ref{fig:double-Fock}, though this
effect is diminished for higher $N$.

\subsection{Nonlinear case: BEC ground state}

We now examine how to enhance the entanglement over the linear case
above, by using a local number-conserving nonlinearity.

We solve for the ground state of a two-component BEC (Fig. \ref{fig:system-two modes}),
as modeled by the following two-mode Hamiltonian \cite{gerdan,Leeprl ,wellbecmurr,esteve}:
\begin{equation}
H=\kappa(a^{\dagger}b+ab^{\dagger})+\frac{g}{2}[a^{\dagger}a^{\dagger}aa+b^{\dagger}b^{\dagger}bb].\label{hamgs-1-1}
\end{equation}
 Here $\kappa$ denotes the conversion rate between the two components,
denoted by the mode operators $a$ and $b$, and $g\propto a_{3D}$
is the nonlinear self interaction coefficient \cite{gerdan}, proportional
to the three-dimensional S-wave scattering length, $a_{3D}$. The
first term proportional to $\kappa$ describes an exchange of particles
between the two wells (modes) in which total number is conserved.
This term is the linear term equivalent to that for a beam splitter.
The second nonlinear term can be thought of as creating squeezing.
The two-mode Hamiltonian model applies to many systems such as optical
cavity modes or superconducting wave-guides with a nonlinear medium. 

\begin{figure}
\begin{centering}
\includegraphics[width=0.85\columnwidth]{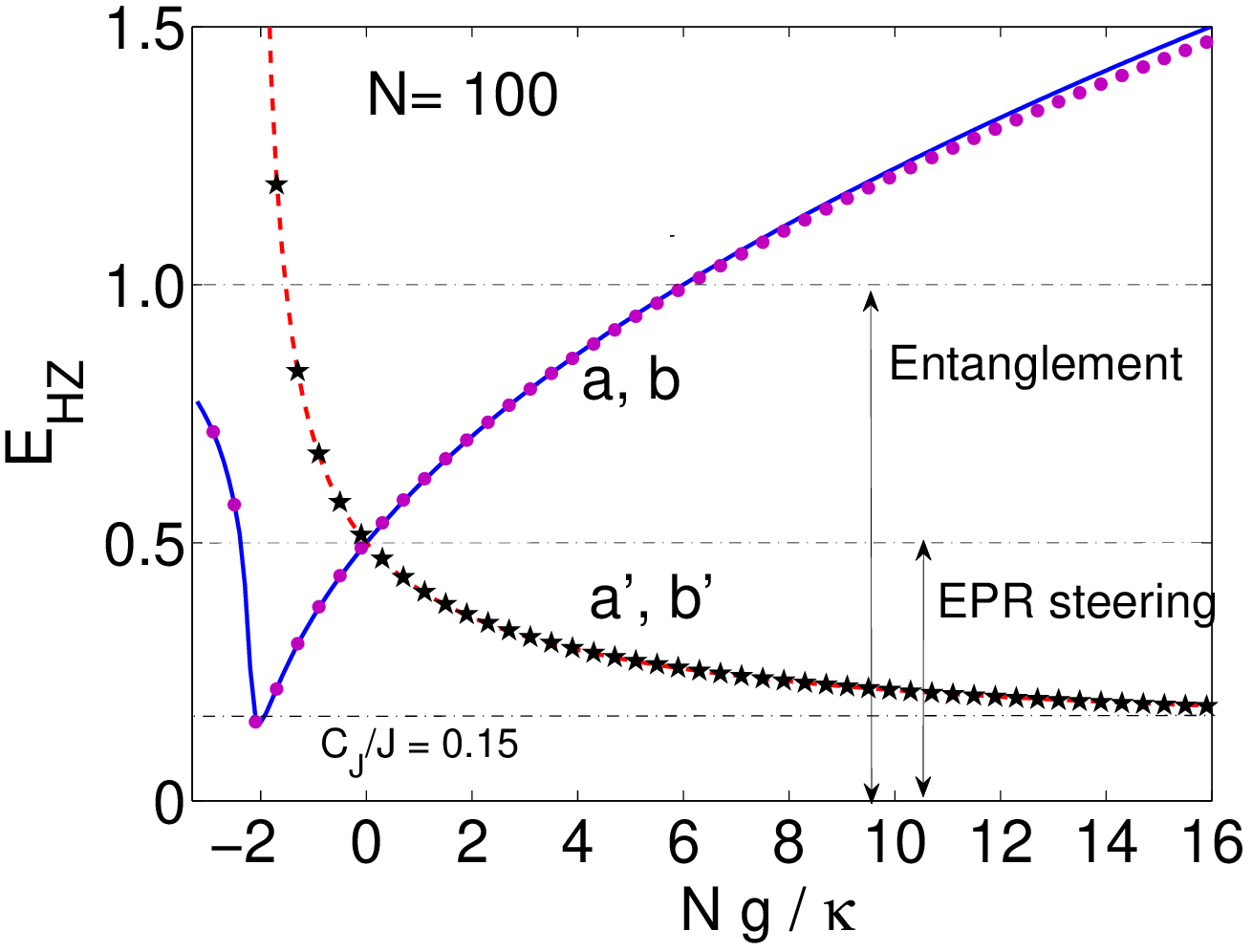} 
\par\end{centering}

\caption{(Color online) Entanglement in the ground state of the BEC Hamiltonian
Eq. (\ref{hamgs-1-1}), using the HZ criterion Eq. (\ref{eq:HZ-spin version-1-1-2}),
plotted against the coupling constant for both positive and negative
couplings, for $N=100$ atoms. Plots show HZ entanglement as a function
of $Ng/\kappa$, with $\kappa>0$ held fixed and $g$ varied. The
mean spin is in the direction defined by $J_{AB}^{X}$. $E_{HZ}<1$
indicates a two-mode entanglement; $E_{HZ}<0.5$ indicates EPR steering.
The dashed red line gives the HZ criterion $E'_{HZ}$ for the rotated
modes $a'$ and $b'$. The predictions for the respective second order
entanglement criterion $E_{HZ}^{(2)}$ (Eq. (\ref{eq:hzsecond-1})),
are given by the dotted and starred curves. \label{fig:sum of two variances-1}}
\end{figure}

The ground state solution is\textcolor{red}{{} }obtained using standard
matrix techniques, and depends only on the dimensionless ratio $g/\kappa$.
\textcolor{black}{We consider a total of $N$ atoms: the number in
well $a$ is $\hat{N_{a}}=a^{\dagger}a$ and in well $b$, $\hat{N_{B}}=b^{\dagger}b$.}

\begin{figure}[h]
\includegraphics[width=0.85\columnwidth]{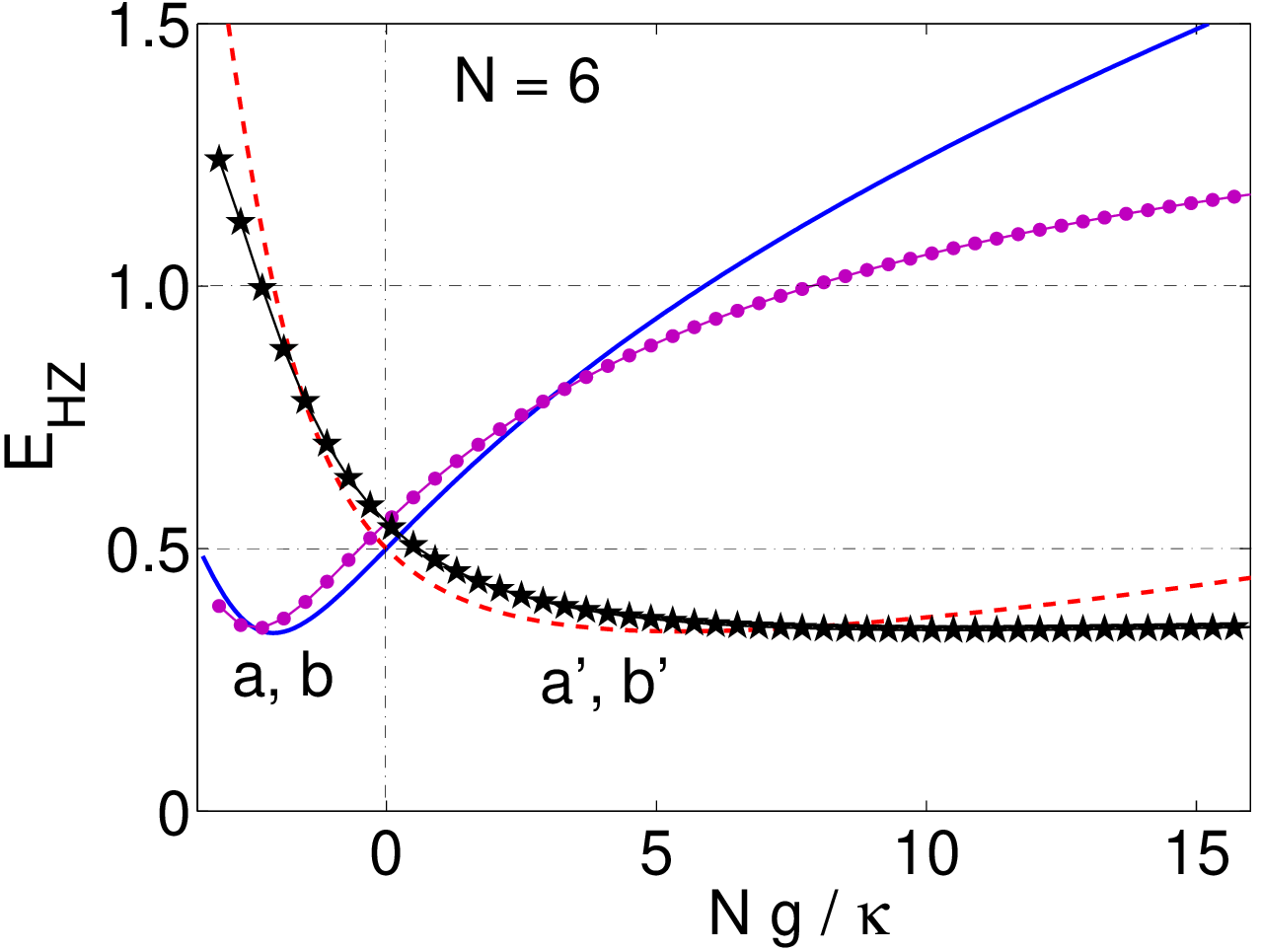}

\caption{(Color online) Same as Fig. \ref{fig:sum of two variances-1} but
for much lower particle number, with $N=6$. First order entanglement
in ($a,b$) is shown by the solid blue line, with second order entanglement
shown by the purple line with dots. First order entanglement in ($a',b'$)
is shown by the dashed red line, with second order entanglement shown
by the black line with stars. The second order entanglement criterion
becomes more sensitive where the nonlinearity is higher in both cases.
\label{fig:Higher-order-entanglement:-1}}
\end{figure}

\textcolor{black}{Solutions show the generation of significant inter-well
two-mode entanglement, including multiparticle entanglement. The entanglement
between the modes $a$ and $b$, and hence between the two wells,
can be detected via the HZ entanglement criterion} Eq. (\ref{eq:hz2}),
for both attractive ($g<0$) and repulsive ($g>0$) regimes. Higher-order
entanglement is also detectable. This result is shown in Figs. \ref{fig:sum of two variances-1}
and \ref{fig:Higher-order-entanglement:-1}.

\subsubsection{Attractive interactions}

\textcolor{black}{The best HZ entanglement (i.e. the smallest possible
value for $E_{HZ}$) is given when the sum of the two variances of
$J_{AB}^{X}$ and $J_{AB}^{Y}$ of (\ref{eq:HZ-spin version-1-1-2})
is minimized. As explained in Section III.D, this sum can never be
zero. }

\textcolor{black}{The best HZ inter-mode entanglement is achieved
in the attractive regime ($g<0$) (as found in $^{41}K$ and $^{7}Li$
isotopes). The absolute lower bound for $E_{HZ}$ is predicted for
the BEC ground state of (\ref{hamgs-1-1}) for a particular critical
value $Ng_{11}/\kappa\approx-2.0$, as shown for $N=100$ in Fig.
\ref{fig:sum of two variances-1}, and for $N=6$ in Fig. \ref{fig:Higher-order-entanglement:-1}.
This critical case has been studied and explained in \cite{cj} and
\cite{phase paper}. We note however that the minimum $E_{HZ}$ becomes
asymptotically small for large $N$. The maximum degree of HZ entanglement
increases with $N$ the number of atoms, according to (\ref{eq:cj-1})
and the relation for $C_{J}$ obtained in \cite{cj}. The degree of
entanglement is strong enough to give EPR steering.}

\textcolor{black}{We note that the strongest theoretical entropic
entanglement $\varepsilon(\rho)$ \cite{entropy,entropy2} is found
for a pure state when all atom numbers are equally represented in
the superposition. It is shown in \cite{bectheoryepr} that the closest
state to this optimum is obtained at a critical value of $Ng_{11}/\kappa\approx-2.0$,
that is, the attractive interaction regime gives rise to a maximal
spread in the distribution of numbers in each well. }

\textcolor{black}{Interestingly, Fig. \ref{fig:sum of two variances-1}
shows that the same point of maximum is observed for the higher order
entanglement measure $E_{HZ}^{(2)}$. This measure can only detect
entanglement that originates from superpositions of the type 
\[
|50\rangle|51\rangle+|51\rangle|50\rangle+|52\rangle|49\rangle
\]
where at least some of the states of the superposition are separated
by $2$ quanta (proved in Section III.B). Similarly, the third order
entanglement criterion $E_{HZ}^{(3)}$ would detect entanglement originating
from states separated by $3$ quanta. In the case of Fig \ref{fig:sum of two variances-1},
where there is $N=100$ quanta, the existence of entangled states
such as $|0\rangle|100\rangle+...+|100\rangle|0\rangle$ could be
detected in principle by measuring $E_{HZ}^{(100)}<1$. This would
give a possible strategy for detecting the entanglement of the NOON
state (the superposition $|N\rangle|0\rangle+|0\rangle|N\rangle$),
though measurement of the higher order moments would present a challenge
\cite{wellbecmurr}. Higher order entanglement (e.g. $E_{HZ}^{(101)}<1$)
would not be possible where the total number of atoms is fixed at
$N$.}

\subsubsection{\textcolor{black}{Repulsive interactions}}

\begin{figure}[h]
\textcolor{black}{\includegraphics[width=0.8\columnwidth]{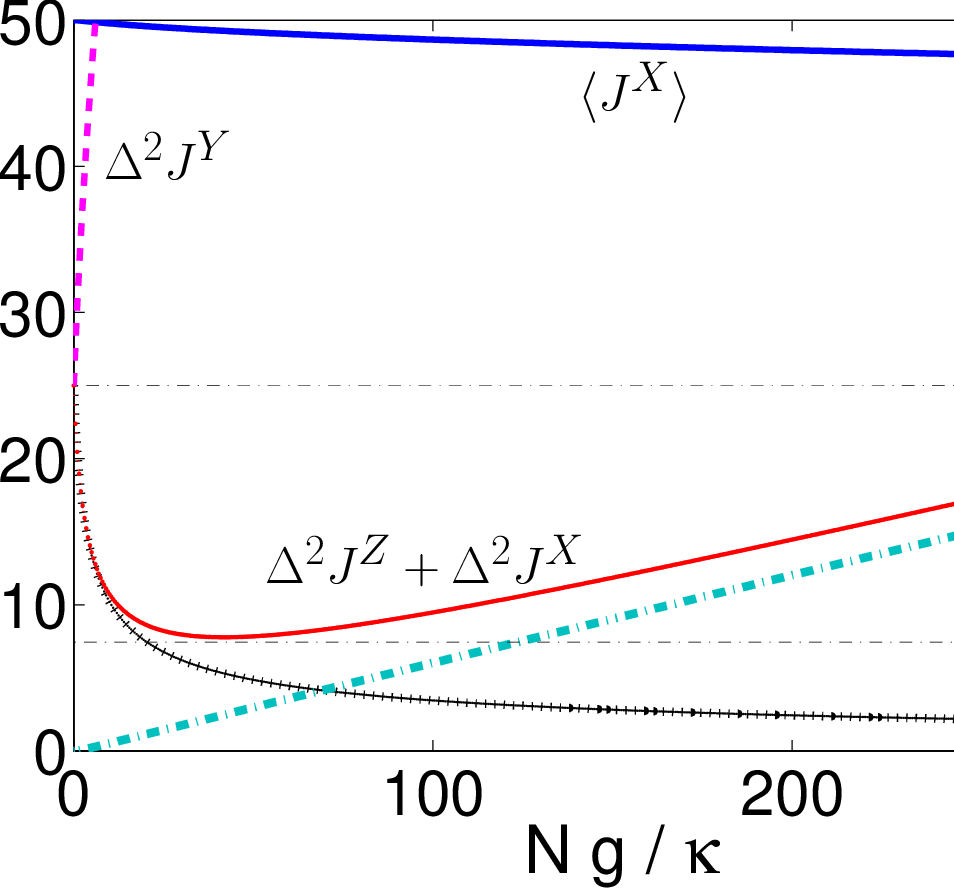}}

\textcolor{black}{\caption{(Color online) The repulsive interaction case for $N=100$, showing
individual spin variances, and mean spin, $\langle J_{AB}^{X}\rangle$,
for the ground state solution of Hamiltonian in the regime where there
is a strong repulsive self-interaction $g/\kappa$, for $N=100$.
Here $\kappa>0$ is fixed and $g$ is varied. \label{fig:3d X-Z plane-1} }
}

\end{figure}

\begin{figure}[h]
 \includegraphics[clip,width=0.5\columnwidth]{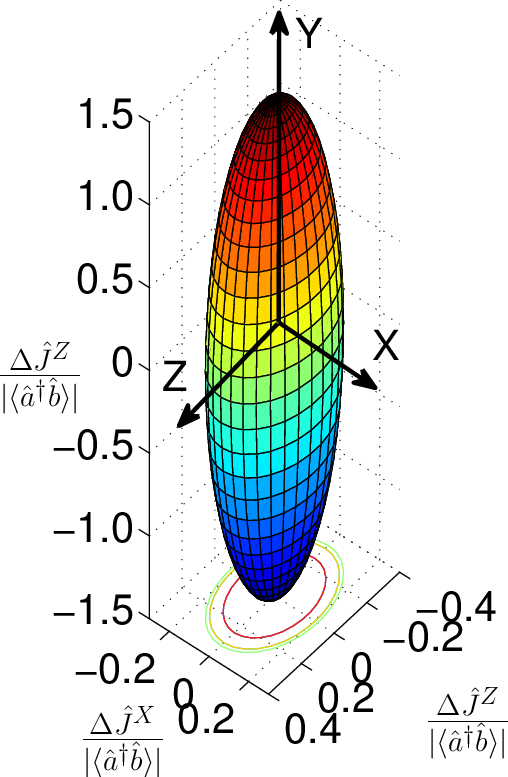}

\caption{(Color online) The 3-D variance ellipsoid corresponding to $N=100$
and a repulsive interaction at the optimum coupling of $Ng/\kappa=40$.
Spin variances are reduced in both axes parallel to the $X-Z$ plane,
to show strong, but not perfect, planar quantum squeezing. The variance
increases perpendicular to the squeezing plane, along the $Y$ axis.\textcolor{red}{{}
}\label{fig:3d X-Z ellipse} }
\end{figure}

\textcolor{black}{The repulsive regime of positive $g$ also predicts
considerable planar squeezing and hence entanglement (Fig. \ref{fig:sum of two variances-1}),
but, in that case, the best planar squeezing is rotated into the $X-Z$
plane as graphed in Fig. \ref{fig:3d X-Z plane-1} \cite{phase paper,guangzhou}.
A depiction of the resulting planar squeezing ellipsoid is shown in
Fig \ref{fig:3d X-Z ellipse}.}

Thus, the corresponding HZ entanglement is between the modes defined
by the\emph{ rotated }coordinates, 
\begin{equation}
a'=\left(a+b\right)/\left(\sqrt{2}i\right),\ b'=\left(a-b\right)/\sqrt{2}.\label{eq:rotated modes}
\end{equation}

The corresponding entanglement criterion is given by: 
\begin{equation}
0<E'_{HZ}=\frac{\left(\Delta J_{AB}^{X}\right)^{2}+\left(\Delta J_{AB}^{Z}\right)^{2}}{\left(\langle a^{\dagger}a\rangle+\langle b^{\dagger}b\rangle\right)/2}\,.
\end{equation}
 The detection of spatial HZ entanglement between the two wells in
the repulsive case would therefore require a different detection scheme,
as proposed in \cite{phase paper}. We note that in both repulsive
and attractive cases, the HZ entanglement can be very significant,
so that the EPR steering nonlocality Eq (\ref{eq:eprsteerhz}) is
predicted via measurement of both the first and second order HZ moments.
Figure \ref{fig:3d X-Z plane-1} indicates that, for fixed $N$, the
repulsive case shows an increasing and then reducing first order HZ
entanglement (\ref{eq:hz2}), as the nonlinearity $g/\kappa$ increases.
The optimum case\textcolor{black}{{} for $N=100$ and a rep}ulsive interaction
occurs at a coupling of $Ng/\kappa=40$. The squeezing ellipsoid for
this coupling is shown in Fig \ref{fig:3d X-Z ellipse}.

\begin{figure}[h]
\includegraphics[width=0.9\columnwidth]{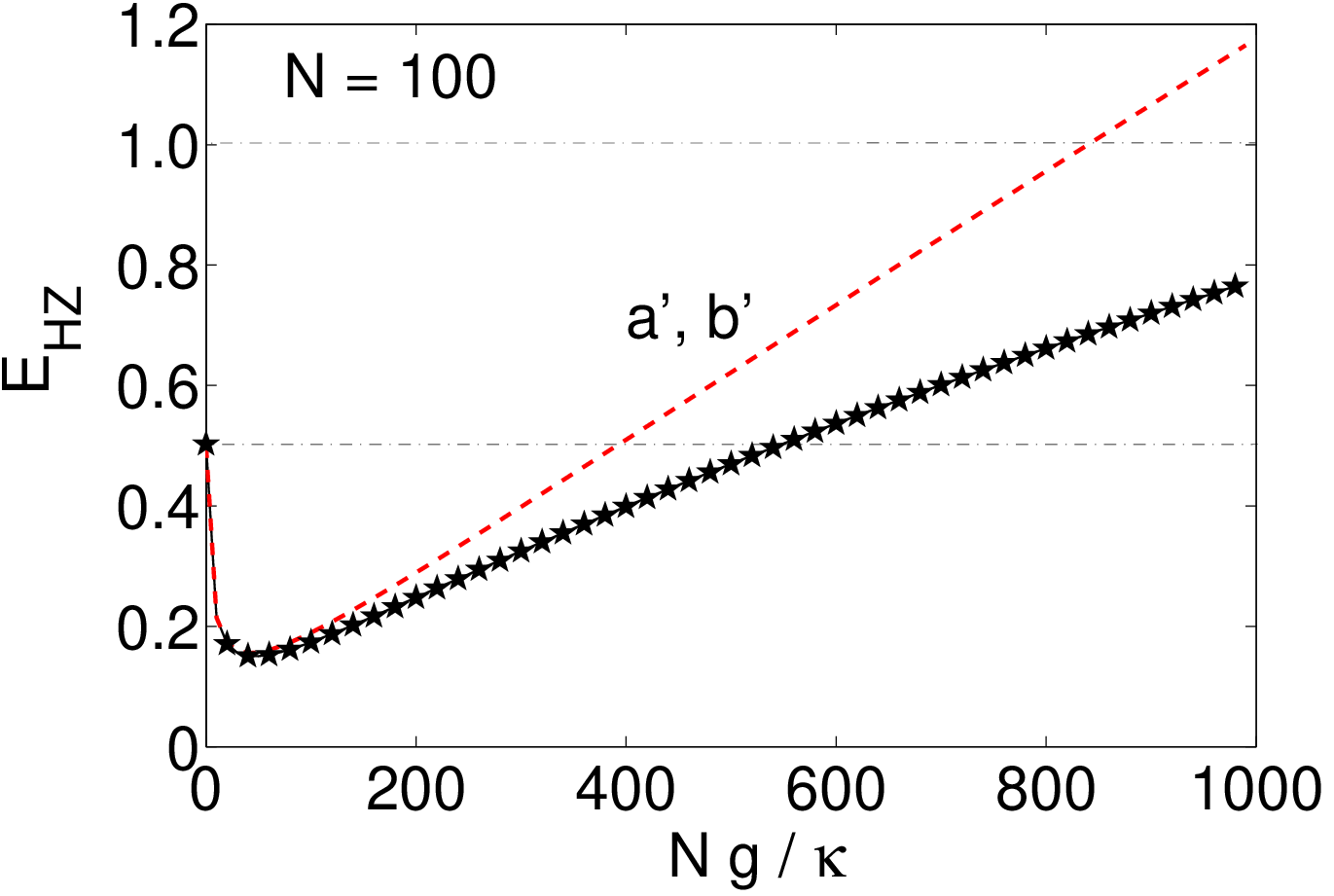}

\caption{(Color online) Higher order entanglement for the case of $N=100$.
Other parameters as in \ref{fig:sum of two variances-1}. The red
dashed line shows a reduced $E_{HZ}$ entanglement as $g/\kappa$
is increased above a certain level. The solid black line (starred)
shows the second order entanglement. Here $E_{HZ}^{(2)}<1$, indicating
entanglement, is possible at higher $g/\kappa$ ratios where the first
order (dashed) criterion shows no entanglement. \label{fig:Higher-order-entanglement:}}
\end{figure}

Interestingly, however, from Fig. \ref{fig:Higher-order-entanglement:},
we see that the second order entanglement criterion for $N=100$ picks
up more entanglement, suggestive that the drop in the entanglement
measured by the first order criterion as the nonlinearity increases
is due to a change in the nature of the entanglement $-$ that it
involves superpositions of states with a greater number difference,
as described in Section III.B $-$ rather than to a loss of entanglement
itself. Fig. \ref{fig:Higher-order-entanglement:-1} shows that a
similar behavior occurs at much lower particle numbers ($N=6$), although
with less overall entanglement at the optimum coupling. In short,
multiparticle entanglement is predicted detectable in the repulsive
case for a wide range of parameter regimes.

\textcolor{black}{We note that a second type of multiparticle entanglement
can be inferred from the degree of first order entanglement. This
approach was proposed in Ref. \cite{sorsolm} and has been used to
infer multiparticle entanglement in Bose Einstein condensates \cite{eprenthiedel,Philipp2010},
based on measurements of the variance of $J_{AB}^{Z}$. This second
type of multiparticle entanglement puts a constraint on the total
number of particles in the entangled state, but can include states
such as 
\[
\{|50\rangle|50\rangle+|49\rangle|51\rangle\}/\sqrt{2}
\]
and is therefore different to that inferred from the higher order
entanglement criteria involving $E_{HZ}^{(m)}$. Where the multiparticle
entanglement is inferred from the first order variances, it is possible
that the states making up the entanglement differ by only one particle
number for each mode. More details for the HZ criterion will be given
in a future paper. }

\subsubsection{Comment on measurement schemes}

The spatial inter-well entanglement can be confirmed, via $E_{HZ}$,
from the measurements of the combined spins $J_{AB}$, using interference
measurements between the two condensates, as has been performed in
\cite{esteve}. Results obtained in this fashion are important in
confirming the existence of entanglement within quantum theory, but
as the measurements are not localized at each site, they cannot be
viewed as rigorous tests of EPR entanglement, steering or nonlocality.
In order to use the above strategies to confirm an EPR-type entanglement,
one would measure the local EPR observables, $X_{A/B}$ and $P_{A/B}$,
at each well \cite{olsenferris,neweprbec}. This is because the moments
of (\ref{eq:hilzub2-1-1-1}) are in terms of operators, $a$ and $b$,
which are linear combinations of the hermitian observables, $X's$
and $P's$. Optically, the $X$ and $P$ are measured using phase
sensitive local oscillators \cite{ou}.

\section{\label{sec:EPR-entanglement:-Four}EPR entanglement: Four component
case}

We examine in this section how to use two additional modes per site
to perform an effective {}``local oscillator'' measurement in this
BEC case. Such strategies have been suggested by Ferris et al \cite{olsenferris}.

\subsection{Linear multimode case }

\begin{figure}[h]

\begin{centering}
\includegraphics[width=0.6\columnwidth]{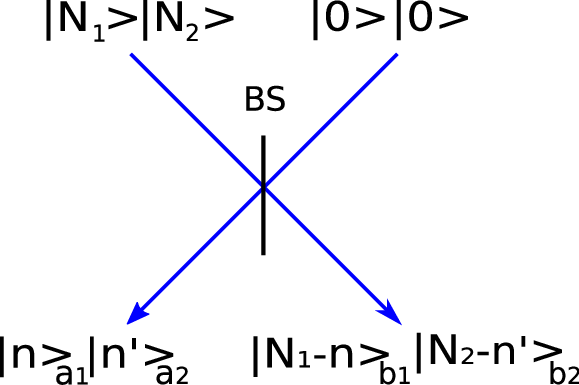} 
\par\end{centering}

\caption{(Color online) We consider pairs of Fock states transmitted through
a beam splitter. Pair $a_{1}$, $b_{1}$ are coupled and become entangled,
as do $a_{2}$, $b_{2}$. \label{fig:model base on Fock state} }
\end{figure}

\begin{figure}[h]

\begin{centering}
\includegraphics[width=0.95\columnwidth]{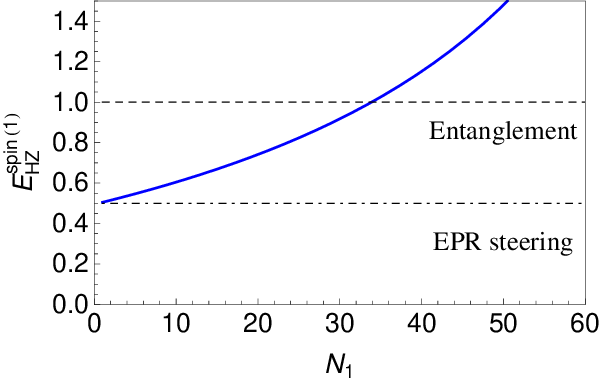} 
\par\end{centering}

\caption{(Color online) The entanglement of pairs of Fock states transmitted
through a beam splitter can be detected via criterion Eq. (\ref{eq:hzspinvarcrit})
for the asymmetric case where pair $a_{2}$ and $b_{2}$ have much
greater numbers $N_{2}\gg N_{1}$ ($N_{2}=100$). Entanglement is
confirmed if $E_{HZ}^{spin\,(1)}<1$. \label{fig:model base on Fock state-1} }
\end{figure}

We study the linear case first, to model a fixed number of atoms with
a minimal BEC nonlinear self interaction. Suppose a Fock number state
$|\psi_{in}\rangle=|N_{1}\rangle_{a_{in1}}|N_{2}\rangle_{a_{in2}}|0\rangle_{b_{in1}}|0\rangle_{b_{in2}}$
is incident on a beam splitter (Fig. \ref{fig:model base on Fock state}),
so that $N_{1}$ and $N_{2}$ are fixed, and modes within each pair
$a_{1},\ b_{1}$ and $a_{2},\ b_{2}$ are coupled by the BS interaction,
with $a_{1}$ and $a_{2}$ (and $b_{1}$, $b_{2}$) remaining uncoupled.
Output modes $a_{1}$ and $b_{1}$ are number-conserved according
to (\ref{eq:sq1-1}); as is pair $a_{2}$, $b_{2}$, and are given
as $a_{1,2}=(a_{in1,2}+b_{in1,2})/\sqrt{2}$, and $ $$b_{1,2}=(a_{in1,2}-b_{in1,2})/\sqrt{2}$.
The output state is 
\begin{eqnarray}
|out\rangle & = & \sum_{n=0}^{N_{1}}\sum_{n'=0}^{N_{2}}c_{n,n'}|n\rangle_{a1}|n'\rangle_{a2}|N_{1}-n\rangle_{b1}|N_{2}-n'\rangle_{b2}\nonumber \\
\end{eqnarray}
 where $c_{n,n'}=\sqrt{N_{1}!N_{2}!}/\sqrt{2^{N_{1}+N_{2}}n!(N_{1}-n)!n'!(N_{2}-n')!}$.
We can evaluate moments, to obtain the prediction for the HZ spin
criterion Eq. (\ref{eq:hzspinvarcrit}). \textcolor{black}{Fig. \ref{fig:model base on Fock state-1}
shows the result of varying $N_{1}$ for fixed $N_{2}=100$. The asymmetric
case is favorable to detecting entanglement.}

\textcolor{black}{Where the initial state is more complex, such as
$|\psi_{in}\rangle=|N_{1}\rangle_{a_{in1}}|N_{2}\rangle_{a_{in2}}|N_{1}\rangle_{b_{in1}}|N_{2}\rangle_{b_{in2}}$,
the output state will }involve superpositions of only even numbers
of atoms in the symmetric and antisymmetric modes, so that $\mid\langle J_{A}^{+}J_{B}^{-}\rangle\mid^{2}=\mid\langle a_{2}^{\dagger}a_{1}b_{2}b_{1}^{\dagger}\rangle\mid^{2}=0$.
As in the case of Section IV.2, we would detect this entanglement
using an appropriate second order spin criterion.

\subsection{Nonlinear four component BEC case}

We now consider the EPR entanglemen\textcolor{black}{t that can be
generated and measured when the modes interact to form the four-mode
BEC ground state. We focus on set-ups that will enable the four mode
case to produce an EPR entanglement that is the replica of the two-mode
HZ entanglement, as displayed in Figures (\ref{fig:sum of two variances-1}-\ref{fig:Higher-order-entanglement:}).
In this case, the second mode at each site may be thought of as part
of a measurement system (Fig. 2).}

\subsubsection{Four-mode BEC Hamiltonian}

\textcolor{black}{We assume the two-well, four-mode system of Fig
\ref{fig:four-modes} is described by the Hamiltonian \cite{fourmodecholsen}:
\begin{equation}
\hat{H}/\hbar=\sum_{i}\kappa_{i}a_{i}^{\dagger}b_{i}+\frac{1}{2}\left[\sum_{ij}g_{ij}a_{i}^{\dagger}a_{j}^{\dagger}a_{j}a_{i}\right]+\left\{ a_{i}\leftrightarrow b_{i}\right\} \,.\label{eq:Hamiltonian}
\end{equation}
 We solve for the ground state of this Hamiltonian. We consider two
modes at each EPR site $A$ and $B$, with four modes in total, as
shown schematically in Fig. \ref{fig:four-modes}. This corresponds
to the two component per well experiments of \cite{Gross2010}, and
somewhat less closely to the multi-mode interferometry experiments
of \cite{Egorov}. Depending on the exact configuration, the local
modes at each EPR site can be independent (in which case local cross
couplings $g_{ij}$ are zero ($g_{12}=0$)), or not independent, as
would be the case where the modes are coupled by the BEC self interaction
term, so the couplings cannot be {}``turned off'', }as in the set-up
of \cite{Gross2010}.\textcolor{black}{{} The coupling constant is
proportional to the three-dimensional S-wave scattering length, so
that $g_{ij}\propto a_{ij}$, as in the two-mode case. For example,
a typical value of the S-wave scattering length for $^{87}Rb$ is
$a_{11}=100.4a_{0}$, where $a_{0}$ is a Bohr radius. Zero cross
couplings are likely to require spatial separation of the two local
modes, as might be achievable with four wells. The quantum dynamics
of the four-well Bose Hubbard model has been studied recently with
two different tunnelling rates \cite{fourmodecholsen}.}

\textcolor{black}{The Hamiltonian (\ref{eq:Hamiltonian}) with $\kappa=\kappa_{1}=\kappa_{2}$
is based on the assumption that the second pair of modes $a_{2},b_{2}$
are coupled between the wells in the same way as the first pair $a_{1},b_{1}$,
which implies similar diffusion across wells. The case where $\kappa_{2}=0$,
$\kappa_{1}\neq0$ is possible where diffusion across the wells can
be controlled, as where the local modes represent separate wells.
We will examine the predictions for both cases.}

\subsubsection{Symmetric tunneling case}

\begin{figure}
\begin{centering}
\textcolor{blue}{\includegraphics[width=0.9\columnwidth]{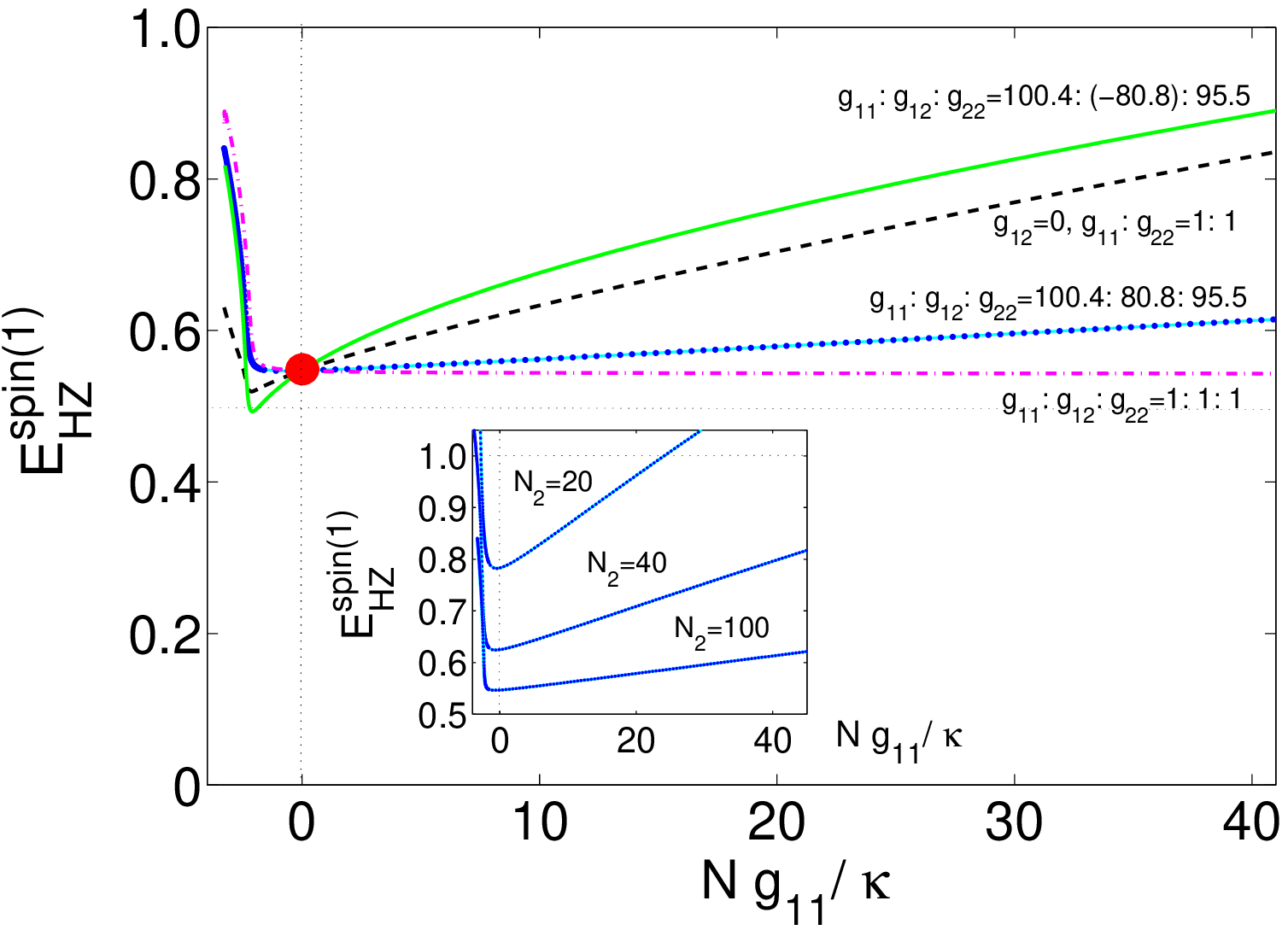}} 
\par\end{centering}

\caption{(Color online) Entanglement using adiabatic cooling to ground state
in a two-well potential, at $T=0K$, for the four mode model of Fig.
\ref{fig:four-modes}, with a variety of local cross couplings $g_{ij}$.
Here $\kappa=\kappa_{1}=\kappa_{2}>0$ and $g_{11}$ is varied with
the other values of $g_{ij}$ held in a fixed ratio, with $N_{1}=5,\ N_{2}=100$.
$E_{HZ}^{spin\,(1)}<1$ indicates entanglement; $ $$E_{HZ}^{spin\,(1)}<0.5$
indicates EPR steering. Curves are labelled in order of nesting as:
(purple dash dotted) equal couplings; (blue dots) non-zero cross-couplings
corresponding to $^{87}Rb$ Feshbach resonance with $a_{11}=100.4a_{0},\ a_{12}=80.8a_{0},\ a_{22}=95.5a_{0}$;
(black dashed) without cross-correlations $g_{12}=0$; $g_{22}=g_{11}$;
(green solid curve) negative relative cross-coupling $g_{11}.g_{12}<0$.
The inset shows the effect of increasingly symmetric atom numbers.\label{fig:four-modes-BEC-1-1} }
\end{figure}

\begin{figure}
\begin{centering}
\textcolor{blue}{\includegraphics[width=0.9\columnwidth]{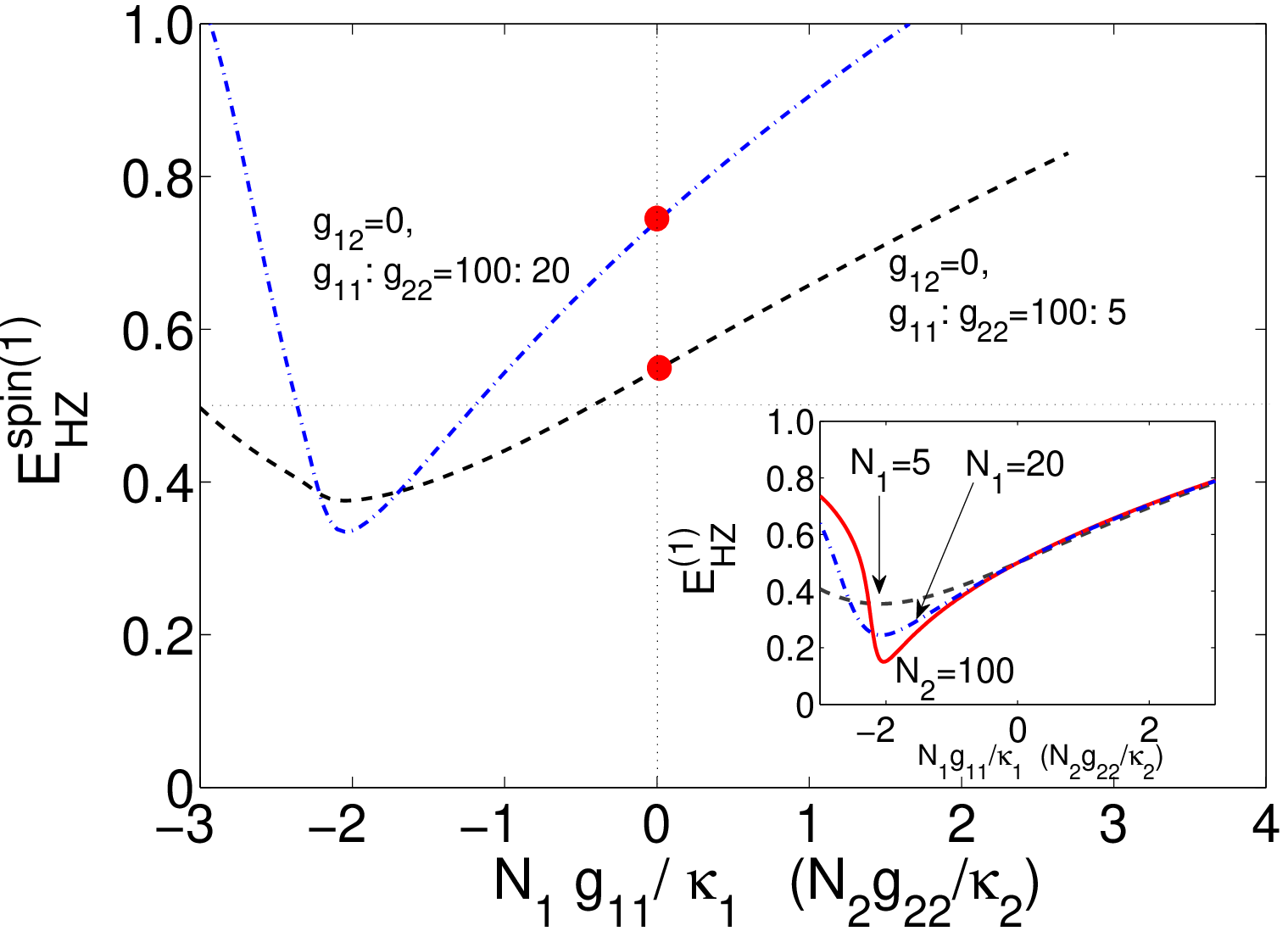}} 
\par\end{centering}

\caption{Entanglement using adiabatic cooling to ground state in a two-well
potential, at $T=0K$. Here \textcolor{black}{$\kappa=\kappa_{1}=\kappa_{2}$},\textcolor{black}{{}
$g_{12}=0$, and both $g_{11}$ and $g_{22}$ are varied so that $N_{1}g_{11}/\kappa_{1}=N_{2}g_{22}/\kappa_{2}$
.} $E_{HZ}^{spin\,(1)}<1$ indicates entanglement; $E_{HZ}^{spin\,(1)}<0.5$
indicates EPR steering. Main graph: (black dashed curve) $N_{1}=5,\ N_{2}=100$;
(blue dotted curve) $N_{1}=20$, $N_{2}=100$. \textcolor{black}{The
curves are for values of local coupling that optimize $E_{HZ}^{(1)}$
for each mode pair, in which case for $N_{2}\gg N_{1}$ the }$E_{HZ}^{spin\,(1)}$
\textcolor{black}{becomes the $E_{HZ}^{(1)}$ displayed in Fig. \ref{fig:sum of two variances-1}.
The inset reveals the individual degree of HZ entanglement $E_{HZ}^{(1)}$
for the mode pairs $a_{1}$,$b_{1}$ and $a_{2}$,$b_{2}$}, as explained
in the text. \textcolor{red}{\label{fig:four-modes-BEC} }}
\end{figure}

\begin{figure}
\begin{centering}
\includegraphics[width=0.92\columnwidth]{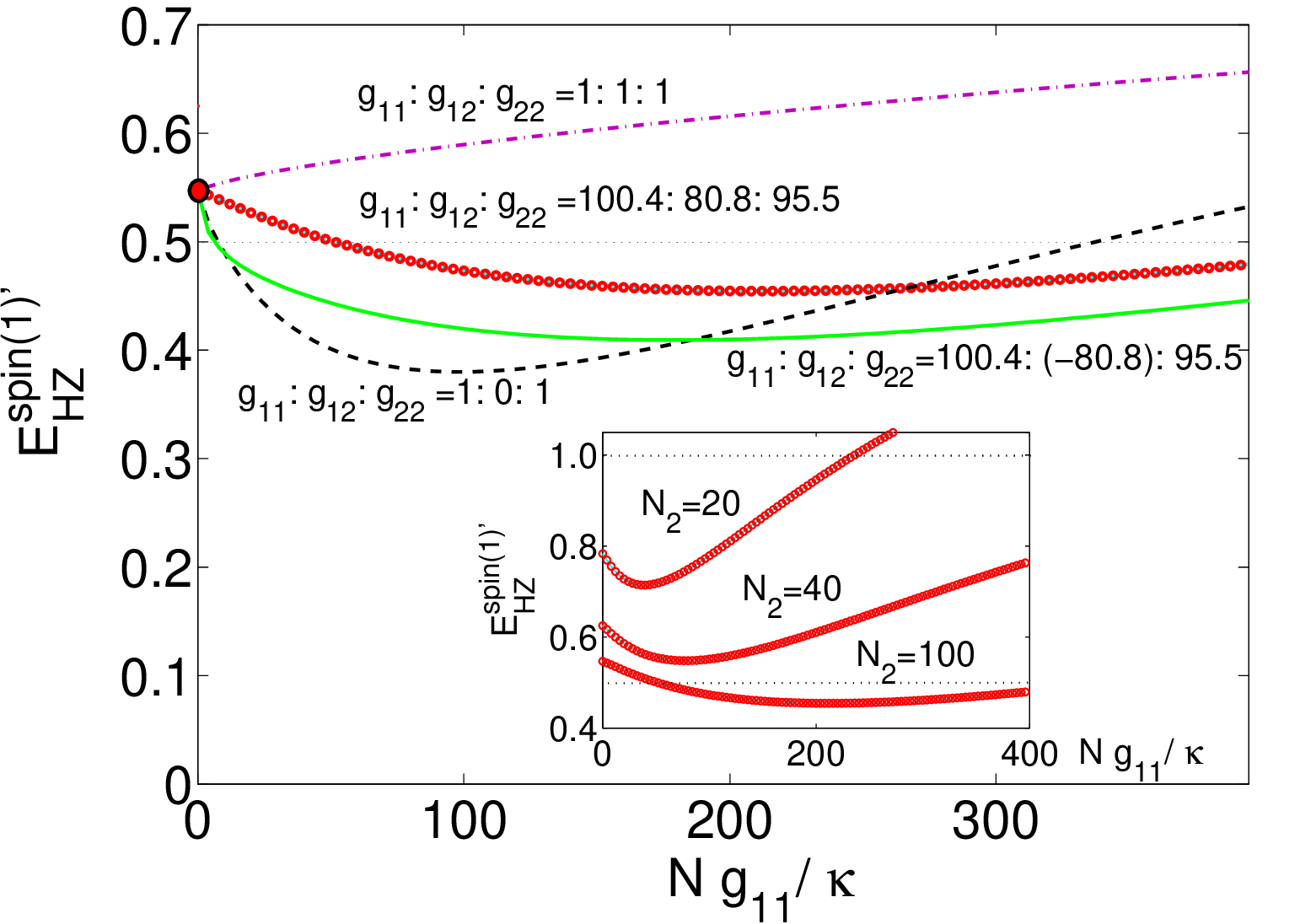} 
\par\end{centering}

\caption{(Color online) Adiabatic cooling to ground state in a two-well potential,
at $T=0K$. Here \textcolor{black}{$\kappa=\kappa_{1}=\kappa_{2}$}.
Parameters are as for Fig. \ref{fig:four-modes-BEC-1-1}, but the
entanglement parameter is calculated for the rotated mode\textcolor{black}{s
$a',\ b'$ of Eq. (\ref{eq:rotated modes}). For large $N_{2}$, the
strength of entanglement measure is enough to confirm EPR steering
via the criterion Eq. (\ref{eq:spinhzeprcritvar})}.\textcolor{blue}{{}
}\label{fig:four-modes-BEC-1}\textcolor{red}{{} }}
\end{figure}

The BEC nonlinearity can enhance the entanglemen\textcolor{black}{t.
This is evident on comparing with the case of zero atom-atom interaction
($g_{ij}=0$), which corresponds to the result of the linear beamsplitter
model (Fig. \ref{fig:model base on Fock state}), and is indicated
by the large red circles in the Figures \ref{fig:four-modes-BEC-1-1}-\ref{fig:four-modes-BEC-1}.
First, we examine the case of symmetric inter-well tunneling with
$\kappa=\kappa_{1}=\kappa_{2}$, so there is complete symmetry between
the nonlocal setups, but a variable loc}al cross coupling $g_{12}$.
Figure \ref{fig:four-modes-BEC-1-1} shows\textcolor{black}{{} entanglement
using the HZ spin criterion }Eq. (\ref{eq:hzspinvarcrit})\textcolor{black}{,
for the ground state, for cases of both zero and strong local couplings
$g_{12}$. Asymmetric atom numbers with $N_{1}\ll N_{2}$ are required
for the best entanglement, however, as shown in the inset of Fig.
}\ref{fig:four-modes-BEC-1-1}\textcolor{black}{. }

\textcolor{black}{We note from Fig. \ref{fig:four-modes-BEC-1-1}
that the entanglement is improved by using a {}``local oscillator''-type
approach, in which the second modes $a_{2}$, $b_{2}$ are independent
of the first at each location ($g_{12}=0$) (being only combined at
the spin measurement stage (\ref{eq:Intrawell-spin})) and are of
much greater numbers ($N_{2}\gg N_{1}$) \cite{olsenferris,eprenthiedel}.
In addition however, we note from the black dashed curve of Fig. \ref{fig:four-modes-BEC}
that better entanglement is obtained if the second {}``local oscillator''
pair $a_{2}$, $b_{2}$ are }\textcolor{black}{\emph{also}}\textcolor{black}{{}
entangled optimally, as given by the critical point of the plots in
Fig. \ref{fig:sum of two variances-1}. }Thus, the optimal $E_{HZ}^{(1)}$
is at \textcolor{black}{$N_{2}g_{22}/\kappa_{2}\approx-2.03$} for
the modes $a_{2}$ and $b_{2}$, and at \textcolor{blue}{${\color{black}N_{1}g_{11}/\kappa_{1}\approx-2.1}$}
for modes $a_{1}$ and $b_{1}$ (as shown in the inset of Fig. \ref{fig:four-modes-BEC}).
The choice $N_{2}g_{22}\sim N_{1}g_{11}$ therefore gives enhanced
EPR spin entanglement (red solid curve of Fig. \ref{fig:four-modes-BEC-1-1}).

The minimum of $E_{HZ}^{spin\,(1)}$ corresponds to the minimum achievable
for the HZ entanglement $E_{HZ}^{(1)}$; this minimum is presented
for the case $N_{1}=100$ in Fig. \ref{fig:sum of two variances-1}.
Better entanglement is thus achieved by increasing the number of atoms
$N_{1}$, provided the other constraints, that $N_{2}\gg N_{1}$ and
$g_{11}$ and $g_{22}$ correspond to the critical choice for each
mode pair, are satisfied, as shown in Fig. \ref{fig:four-modes-BEC}.\textcolor{blue}{{}
}\textcolor{black}{Analytical details are given in the Appendix.}

\textcolor{black}{It is interesting that the case of approximately
equal couplings $g_{11}=g_{22}=g_{12}$ is generally less favorable
for the HZ spin entanglement (Figure \ref{fig:four-modes-BEC-1-1}).
This can be understood if we rewrite the Hamiltonian (\ref{eq:Hamiltonian})
in terms of the spin operators. }We obtain $H\simeq\chi\left(J_{A}^{Z}\right)^{2}+\chi\left(J_{B}^{Z}\right)^{2}+\kappa(a_{1}^{\dagger}b_{1}+a_{1}b_{1}^{\dagger}+a_{2}^{\dagger}b_{2}+a_{2}b_{2}^{\dagger})$,
where $\chi\simeq\frac{1}{2}(g_{11}+g_{22}-2g_{12})$ gives the effective
nonlinearity, and those terms related to $J_{A,B}^{Z},\ N_{1,2}^{2},\ N_{1,2}$
have been omitted.\textcolor{black}{{} For equal couplings $g_{12}=g_{11}=g_{22}$,
the Hamiltonian thus effectively reduces to the linear term of the
BS model of Fig. \ref{fig:model base on Fock state}, the predictions
of which are given by the red circles in Figs 14 and 15. This is evident
in the results of Figs. \ref{fig:four-modes-BEC-1-1} and \ref{fig:four-modes-BEC}.
Furthermore, enhancement of the nonlinearity is possible, if $g_{12}$
becomes negative. The green solid curve of Figure \ref{fig:four-modes-BEC-1-1}
shows an enhanced entanglement for negative local cross-coupling,
$g_{12}<0$. }

\textcolor{black}{As is consistent with the two-mode results, the
spin HZ entanglement is optimal in the attractive regime, $g_{11}<0$.
Enhancement of entanglement in the repulsive regime is possible (Fig.
\ref{fig:four-modes-BEC-1}), if one examines the spin HZ entanglement
for the rotated modes, $a',\ b'$ of Eq. (\ref{eq:rotated modes}).}

\begin{figure}
\centering{}\includegraphics[width=0.9\columnwidth]{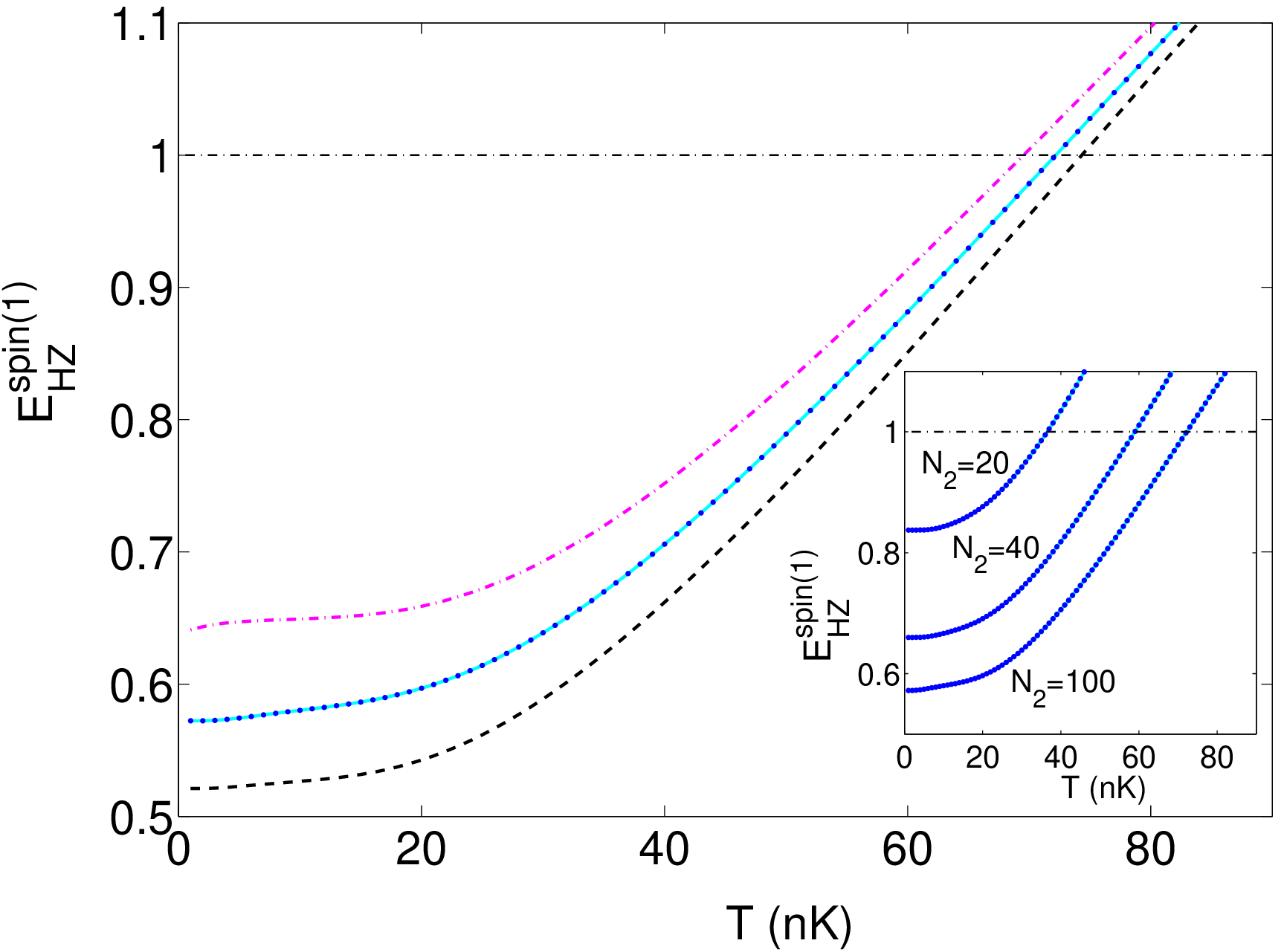}\caption{(Color online) \textcolor{black}{The effect of critical temperatures
corresponding to the parameters of Fig. }\ref{fig:four-modes-BEC-1-1}\textcolor{black}{{}
when $Ng/\kappa\approx-2.23$. \label{fig:critical T}} }
\end{figure}

\textcolor{black}{The effect of temperature is presented in Fig. \ref{fig:critical T}.
In our calculations, we account for finite temperatures by assuming
a canonical ensemble of $\rho=\exp[-H/k_{B}T]$, with an inter-well
coupling of $\kappa/k_{B}=50nK$. The critical temperature for the
spin HZ entanglement signature is shown in Fig. \ref{fig:critical T}. }

\subsubsection{Asymmetric tunneling case}

\begin{figure}[h]
 \centering{}\includegraphics[width=0.9\columnwidth]{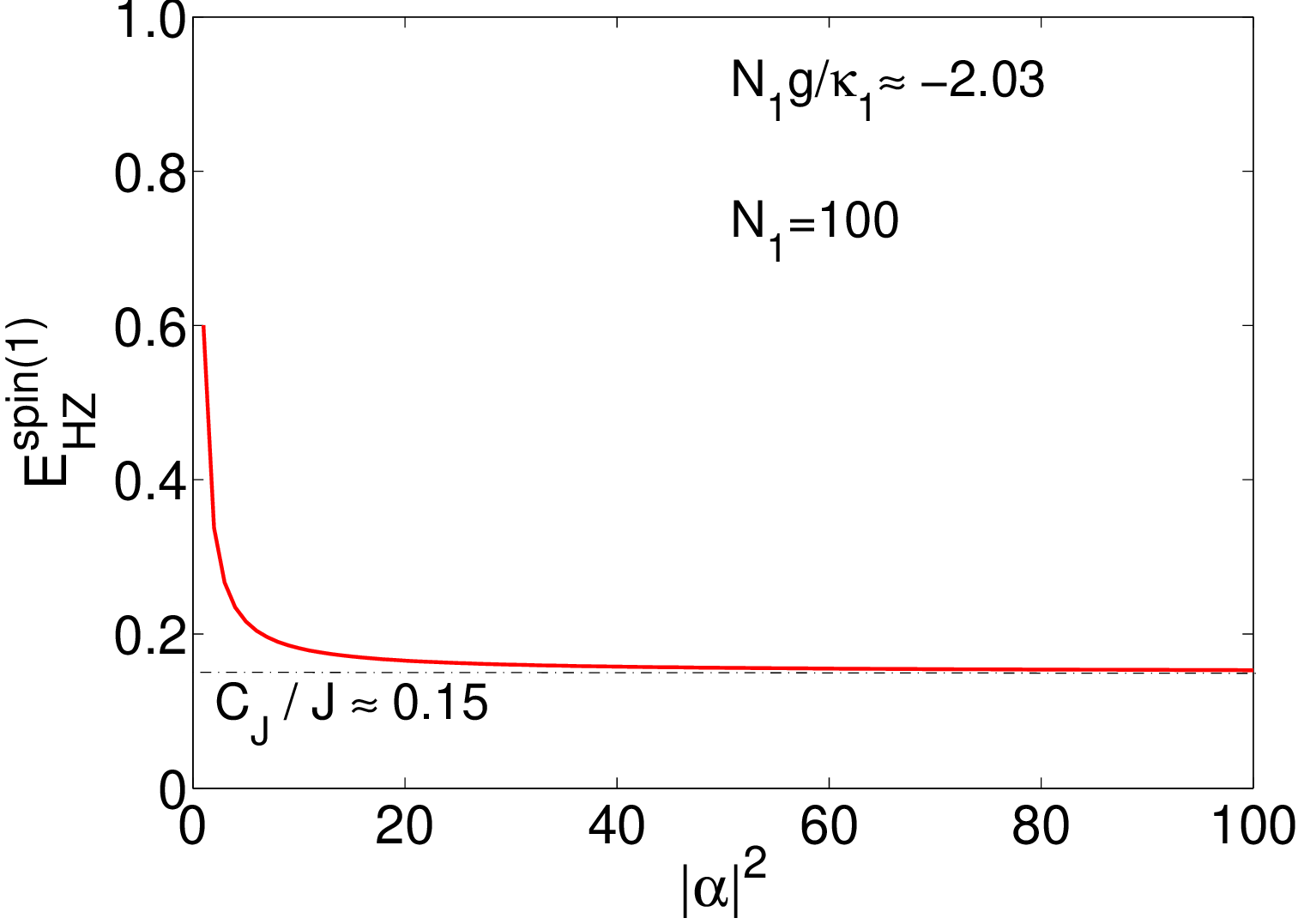}\caption{(Color online) \textcolor{black}{The effect of an uncorrelated coherent
atomic oscillator field for mode $2$ in a coherent state with amplitude
$\alpha$, in the optimal case of Fig. \ref{fig:sum of two variances-1}
when $N_{1}g/\kappa\approx-2.03$. \label{fig:k2=00003D00003D0}} }
\end{figure}

An alternative strategy more closely aligned to those used in optics
is to consider $\kappa_{2}=0$ , $\kappa_{1}\neq0$. In this case,
the modes $a_{2}$ and $b_{2}$ are uncoupled and independent. If
they are prepared in coherent states $|\alpha_{2}\rangle|\beta_{2}\rangle$
(we take $\alpha_{2}=\beta_{2}=\alpha$, where $\alpha$ is real),
with $\alpha$ large, the entanglement $E_{HZ}^{spin\,(1)}$ approaches
the value given in the two-mode case, by $E_{HZ}^{(1)}$. We explain
this as follows. For independent modes, as shown by equation (\ref{eq:simindephzspin-1})
of the Appendix, the HZ spin entanglement criterion (\ref{eq:entspinhilzub})
becomes, upon assuming coherent states for $a_{2}$ and $b_{2}$,
\begin{equation}
|\langle a_{1}^{\dagger}b_{1}\rangle|^{2}\alpha^{4}>\langle a_{1}^{\dagger}a_{1}b_{1}^{\dagger}b_{1}\rangle(1+\alpha^{2})^{2}\label{eq:k2=00003D00003D0-1}
\end{equation}
 which we see will approach the required two-mode entanglement level
in the limit of large $\alpha$. \textcolor{black}{Figure \ref{fig:k2=00003D00003D0}
plots the result with finite numbers of atoms for the case of optimal
$E_{HZ}^{(1)}$ which occurs at $N_{1}g_{11}/\kappa_{1}\approx-2.03$
when $N_{1}=100$. We can see that the four mode EPR entanglement
achieved ($C_{J}/J\approx0.15$) is that of the two-mode case (Fig.
\ref{fig:sum of two variances-1}) provided there is a large enough
number of atoms in the second mode.}

\section{Conclusion\label{sec:Summary}}

We have examined strategies capable of generating detectable entanglement
between two spatially-separated potential wells in a BEC. These include
both two and four-mode strategies similar to those already used for
spin-squeezing, but generalized to a double well. The model used to
calculate the relevant variances has been shown to give a good fit
to experimental data \cite{guangzhou,Gross2010}. Our results find
that local cross couplings can have a strong effect on entanglement,
and results for the EPR entanglement improve with higher atom numbers.
We find that a spin version of the Hillery-Zubairy (HZ) entanglement
criterion appears readily suited to analyzing entanglement and the
EPR steering paradox in these experiments\textcolor{black}{. Furthermore,
we have shown that the higher order HZ entanglement criteria can give
information about the number of particles involved in the entangled
state and the nature of the multiparticle entanglement.} 

\textcolor{black}{The predictions in this paper are based on the assumption
that the total number $N$ of atoms is fixed. Entanglement ($E_{HZ}^{(1)}=0.5$),
though not EPR-steering, is obtainable between the output ports of
a beam splitter with a number (Fock) state input, in the absence of
nonlinear coupling terms, as was shown in Sections \ref{sec:generation-of-number-conserving}.A
and \ref{sec:EPR-entanglement:-Four}.A. However, for coherent state
inputs, which have a Poissonian number distribution, this entanglement
is not possible \cite{cohbs}, and we draw the conclusion that number
fluctuations will have an important effect on the entanglement. The
effect of particle fluctuations on entanglement and precision measurement
has been studied recently by Hyllus et al \cite{partiintnumberfluctuation}
and He et al \cite{bectheoryepr,phase paper}. However, we make the
final note that these studies do not treat the EPR steering nonlocality.}
\begin{acknowledgments}
This research was supported by an Australian Research Council Discovery
grant. We wish to acknowledge useful discussions with M. K. Oberthaler,
C. Gross, P. Treutlein, A. I. Sidorov, M. Egorov and B. Opanchuk. 
\end{acknowledgments}

\section*{Appendix }

\textcolor{black}{In this Appendix, we show we show how to directly
{}``convert'' the inter-well entanglement shown in Fig. \ref{fig:sum of two variances-1}
to an EPR entanglement, with the use of a {}``local oscillator''-type
treatment which applies where two of the strong local modes are uncorrelated.
This is the case of $g_{12}=0$, illustrated in Fig. \ref{fig:four-modes}.}

Local oscillator measurements are achieved optically by combining
a mode with a very strong coherent state \cite{ou}. We can achieve
something effectively equivalent to a {}``local oscillator'' measurement,
where the second pair of levels $a_{2}$, $b_{2}$ are much more heavily
populated than levels $a_{1}$ and $b_{1}$, by assuming the second
pair of modes are in an uncorrelated coherent state. We explain this
as follows. Since $J_{A}^{+}=a_{1}^{\dagger}a_{2}$ and $J_{A}^{-}=a_{1}a_{2}^{\dagger}$
and $J_{B}^{+}=b_{1}^{\dagger}b_{2}$ and $J_{B}^{-}=b_{1}b_{2}^{\dagger}$
we can rewrite the criterion (\ref{eq:entspinhilzub}) in terms of
the mode operator moments, for this special case, by the factorization
that is justified for independent fields at each location. Thus, 
\begin{equation}
|\langle J_{A}^{+}J_{B}^{-}\rangle|^{2}=|\langle a_{1}^{\dagger}b_{1}\rangle\langle a_{2}b_{2}^{\dagger}\rangle|^{2}\,,
\end{equation}
 and similarly 
\begin{equation}
\langle(J_{A}^{+}J_{A}^{-})(J_{B}^{+}J_{B}^{-})\rangle=\langle a_{1}^{\dagger}a_{2}a_{1}a_{2}^{\dagger}b_{1}^{\dagger}b_{2}b_{1}b_{2}^{\dagger}\rangle\,.
\end{equation}
 The criterion (\ref{eq:entspinhilzub}) becomes 
\begin{eqnarray}
|\langle a_{1}^{\dagger}b_{1}\rangle|^{2}|\langle a_{2}b_{2}^{\dagger}\rangle|^{2} & > & \langle a_{1}^{\dagger}a_{1}b_{1}^{\dagger}b_{1}\rangle\langle(1+a_{2}^{\dagger}a_{2})(1+b_{2}^{\dagger}b_{2})\rangle\nonumber \\
\label{eq:simindephzspin-1}
\end{eqnarray}

Clearly, since the inter-well entanglement studied in Section \ref{sec:generation-of-number-conserving}
and summarized in Fig. \textcolor{black}{\ref{fig:sum of two variances-1}}
enables $|\langle a_{1}^{\dagger}b_{1}\rangle|^{2}>\langle a_{1}^{\dagger}a_{1}b_{1}^{\dagger}b_{1}\rangle$
via the HZ entanglement criterion, we will have (at least) the same
level of four mode EPR entanglement, provided 
\begin{equation}
|\langle a_{2}b_{2}^{\dagger}\rangle|^{2}\geq\langle(1+a_{2}^{\dagger}a_{2})(1+b_{2}^{\dagger}b_{2})\rangle.\label{eq:loassumption-1}
\end{equation}
 \textcolor{red}{{} }In fact, the inequality would represent violation
of the two-site version of the Bell inequality discussed in \cite{CFRD},
which is not achievable for this system. However, it is still possible
to optimize the EPR entanglement. This can be achieved in the following
way. If the two modes $a_{2}$ and $b_{2}$ are also coupled via an
inter-well interaction ($\kappa_{2}\neq0$\textcolor{black}{{} in
Fig. \ref{fig:four-modes}), to produce the ground state solution
of Fig \ref{fig:3d X-Z plane-1}, then $E_{HZ}^{(1)}<1$ amounts to
$|\langle a_{2}b_{2}^{\dagger}\rangle|^{2}>\langle a_{2}^{\dagger}a_{2}b_{2}^{\dagger}b_{2}\rangle$.
The optimal $E_{HZ}^{(1)}$ is at} \textcolor{black}{$N_{2}g_{22}/\kappa_{2}\approx-2.03$},
while for the modes $a_{1}$ and $a_{2}$, the optimal (\ref{eq:simindephzspin-1})
occurs for \textcolor{blue}{${\color{black}N_{1}g_{11}/\kappa_{1}\approx-2.1}$}
(inset of Fig. \ref{fig:four-modes-BEC}). This choice gives enhanced
EPR entanglement as shown in Fig. \ref{fig:four-modes-BEC-1-1}. Better
entanglement is possible for this optimal choice, as the numbers are
increased (Fig. \ref{fig:four-modes-BEC}).\textcolor{blue}{{} }

\end{document}